\documentclass
[superscriptaddress,secnumarabic,amssymb,amsmath,nobibnotes,aps,prd,showpacs,showkeys,nofootinbib,onecolumn]{revtex4}%
\usepackage{graphicx}
\usepackage{subfigure}
\usepackage{epsf}
\usepackage{bm}
\usepackage{amsmath}
\usepackage{amsfonts}
\usepackage{amssymb}
\usepackage{color}%
\providecommand{\U}[1]{\protect\rule{.1in}{.1in}}

\newcommand{\be}{\begin{equation}}
\newcommand{\ee}{\end{equation}}

\newcommand{\mincir}{\raise
-3.truept\hbox{\rlap{\hbox{$\sim$}}\raise4.truept\hbox{$<$}\ }}
\newcommand{\magcir}{\raise
-3.truept\hbox{\rlap{\hbox{$\sim$}}\raise4.truept\hbox{$>$}\ }}

\begin{document}
\title{Interacting quintessence in light of Generalized Uncertainty Principle: Cosmological perturbations and dynamics}
\author{Andronikos Paliathanasis}
\email{anpaliat@phys.uoa.gr}
\affiliation{Institute of Systems Science, Durban University of Technology, Durban 4000,
South Africa}
\affiliation{Instituto de Ciencias F\'{\i}sicas y Matem\'{a}ticas, Universidad Austral de
Chile, Valdivia, Chile}
\author{Genly Leon}
\email{genly.leon@ucn.cl}
\affiliation{Departamento de Matem\'{a}ticas, Universidad Cat\'{o}lica del Norte, Avda.
Angamos 0610, Casilla 1280 Antofagasta, Chile}
\author{Wompherdeiki Khyllep}
\email{sjwomkhyllep@gmail.com}
\affiliation{Department of Mathematics, North-Eastern Hill University,
	Shillong, Meghalaya 793022, India}
\affiliation{Department of Mathematics,
	St.\ Anthony's College, Shillong, Meghalaya 793001, India}
\author{Jibitesh Dutta}
\email{jibitesh@nehu.ac.in}
\affiliation{Mathematics Division, Department of Basic Sciences and Social
Sciences, North-Eastern Hill University,  Shillong, Meghalaya 793022, India}
\affiliation{Inter University Centre for Astronomy and Astrophysics, Pune
411 007, India }
\author{Supriya Pan}
\email{supriya.maths@presiuniv.ac.in}
\affiliation{Department of Mathematics, Presidency University, 86/1 College Street, Kolkata 700073, India}

\begin{abstract}

We consider a cosmological scenario endowed with an interaction between the universe's dark components $-$  dark matter and dark energy. Specifically, we assume the dark matter component to be a pressure-less fluid, while the dark energy component is a quintessence scalar field with Lagrangian function modified by the quadratic Generalized Uncertainty Principle. The latter modification introduces new higher-order terms of fourth-derivative due to quantum corrections in the scalar field's equation of motion. Then we investigate asymptotic dynamics and  general behaviour of solutions of the field equations for some interacting models of special interests in the literature. At the background level, the present interacting model exhibits the matter-dominated and de Sitter solutions which are absent in the corresponding quintessence model.
Furthermore, to boost the background analysis,  we study cosmological linear perturbations in the Newtonian gauge where we show how perturbations are modified by quantum corrected terms from the quadratic Generalized Uncertainty Principle. Depending on the coupling parameters, scalar perturbations show a wide range of behavior.

\end{abstract}
\date{\today}
\maketitle

\section{Introduction}
\label{sec1}

The physics of the dark sector of our Universe is obscure, at least according to the up-to-date observational shreds of evidence. This dark sector comprises of a dark matter (DM) which is almost pressure-less or cold, and a dark energy (DE) fluid, and they jointly contribute nearly 96\% ($\sim$ 68\% for DE and $\sim$ 28\% for DM) of the entire energy content of the Universe \cite{Aghanim:2018eyx}. The most mathematically simplest cosmological model that explains this dark sector is the $\Lambda$-Cold Dark Matter ($\Lambda$CDM) in which $\Lambda >0$, the cosmological constant, plays the role of DE. Even though the $\Lambda$CDM model has been found to fit excellently to most of the available astrophysical and cosmological probes, the most significant assumption within such a cosmological picture is the independent evolution of both DM and DE, that means, in other words, we have a non-interacting cosmological scenario. However, despite the tremendous success of the $\Lambda$CDM model, there are already known severe issues associated with it. We can reconcile the existing problems related to the  $\Lambda$CDM model by an alternative description of the Universe along with the excellent fits to the available observational data. The list of the models is quite extensive, including a variety of cosmological theories and models; see for instance \cite{Copeland:2006wr}. Amongst many attractive cosmological models,  in this article, we focus on a very generalized cosmological theory where DM and DE are freely allowed to interact with one other, known as Interacting DE (IDE) or coupled DE.

The IDE theory which was formally proposed by Amendola \cite{Amendola:1999er}  received tremendous attention in the scientific community for its ability to offer a possible explanation to the cosmic coincidence problem \cite{Huey:2004qv,Cai:2004dk,Pavon:2005yx,Berger:2006db,delCampo:2006vv,delCampo:2008sr,delCampo:2008jx} where a direct interaction between DM and DE (in a non-gravitational way) can play the game. However, we recall an earlier work by Wetterich  \cite{Wetterich:1994bg} where the author first argued the possibility of an interaction in the Universe where a coupling or interaction between a scalar field and gravity can lead to a time-dependent cosmological `constant' which asymptotically becomes constant and consequently a possible explanation to the cosmological constant problem was placed. With such motivations, IDE theory was investigated widely in the literature.  Subsequently  it was proved that interaction between DM and DE could offer some more interesting possibilities, for example, the crossing of the phantom divide line without any need to scalar field  \cite{Huey:2004qv,Wang:2005jx, Das:2005yj, Sadjadi:2006qb, Pan:2014afa}, and most importantly, it can explain the recent $H_0$ tension \cite{Kumar:2017dnp, DiValentino:2017iww,Kumar:2019wfs} and the $\sigma_8$ tension \cite{Pourtsidou:2016ico,An:2017crg,Kumar:2019wfs}. We refer to a recent review on the $H_0$ tension where the observational data report that IDE rocks (see Tables B1 and B2 of  \cite{DiValentino:2021izs}).  Nevertheless, the questions regarding the derivation of the interaction function from some fundamental action have not been truly understood yet. Attempts to derive the interaction function have been made by various investigators with some reasonable answers, see for instance \cite{Gleyzes:2015pma,Boehmer:2015kta,Boehmer:2015sha,Pan:2020zza,vandeBruck:2015ida,Xiao:2018jyl,Amico:2016qft,Pan:2020mst}, however, the development of this section is still in progress.

In this article, we consider an interacting scheme where DM is pressure-less, and DE is a scalar field. Even though we are considering an interacting scalar field model which reminds us the earlier works in this direction \cite{Amendola:1999er,Huey:2004qv,Kase:2019veo}, however, the framework and the formulation of the present interacting scalar field scenario is different from them \cite{Amendola:1999er,Huey:2004qv,Kase:2019veo} because here the scalar field sector which is interacting with the pressure-less DM has been modified following the generalized uncertainty principle (GUP). The inclusion of GUP adds a novel feature to this work. According to the past historical records, GUP has many astrophysical and cosmological implications, for instance, the origin of the magnetic fields in the Universe sector \cite{AM},  black hole thermodynamics \cite{ACS} and some others (see \cite{gup1,gup2} for a detailed description). Since GUP plays a vital role in the context of the early physics of the Universe, therefore, its effects on the late Universe physics is a topical issue that should be investigated, aiming to understand the complete dynamical picture of the Universe. 
It is always very appealing to encounter the new physical theories modified by the GUP's inclusion to understand the minimum length scale effects in the late Universe physics. Following this motivation, the authors of \cite{gup2} first investigated the dynamics of the GUP modified quintessence scalar field with two different potentials; one is exponential, and the other is an arbitrary. The resulted dynamics offered some interesting possibilities that are absent in the usual quintessence scalar field model. For instance, the appearance of the critical points describing the de Sitter Universe in the GUP modified quintessence is a new piece of result, and this does not depend on the choice of the potential. In this article, we extended the previous work \cite{gup2} by allowing an interaction between the GUP modified quintessence scalar field and the pressure-less DM to examine the dynamical features of this generalized cosmic scenario.

To improve the background analysis, we also investigate cosmological linear perturbations.   More specifically, we study the dynamics of scalar perturbations during the matter-dominated era, where the effect of GUP might imprint exciting features during the growth of structure. Such analysis is crucial to identify specific signatures of the interacting scalar field theories in the light of GUP to look for by cosmological observations. Therefore, the present study will serve as a preliminary investigation of the modified IDE model's cosmic viability. 

The article is structured as follows:
In Section \ref{sec2} we present the basic equations of the interacting scenario. For the DE, we assume that it is described
by a minimally coupled scalar field where the GUP modifies the Action Integral. 
The new set of equations are presented in Section \ref{sec3}. We
continue our analysis to study the asymptotic solutions for the field equations for a linear interacting model. The detailed analysis of the
asymptotic solutions and their physical properties are given in Section
\ref{sec4}.
Furthermore, in Section \ref{sec5}, we derive the scalar equations
of the cosmological linear perturbations. We find that the perturbed system is also a singular perturbation system due to the new terms introduced by GUP.
Therefore, the qualitative evolution of the perturbed equations in the
slow-fast manifolds is studied in the matter-dominated era from where we find
new growing modes that follow from GUP corrections. Finally, in Section \ref{sec6} we summarize the main results of this article.

\section{Interacting dark energy}
\label{sec2}

According to cosmological principle, our Universe in the large scales is almost homogeneous and isotropic. 
Therefore, we consider that the background physical
space is described by 
Friedmann--Lema\^{\i}tre--Robertson--Walker (FLRW) line-element with zero curvature given by
\begin{equation}
ds^{2}=-dt^{2}+a^{2}(t)\left(  dx^{2}+dy^{2}+dz^{2}\right)  , \label{gp.01}%
\end{equation}
where $a\left(  t\right)  $ is the expansion scale factor and $(t, x, y, z)$ are the co-moving coordinates.
In the context of Einstein's General Relativity (GR) the gravitational field
equations are (in the units where $8 \pi G =1$): 
\begin{equation}
G_{~\nu}^{\mu}=T_{~\nu}^{\mu},\;\;  \mu, \nu=0,~1,~2,~3. \label{gp.02}%
\end{equation}
where $G_{~\nu}^{\mu}=R_{~\nu}^{\mu}-\frac{1}{2}Rg_{~\nu}^{\mu}$ is the
Einstein tensor and $T_{~\nu}^{\mu}$ is the effective momentum tensor%
\begin{equation}
T_{\mu\nu}=T_{\mu\nu}^{\left(  m\right)  }+T_{\mu\nu}^{\left(  d\right)
}+T_{\mu\nu}^{\left(  r\right)  }+T_{\mu\nu}^{\left(  b\right)  },
\label{gp.03}%
\end{equation}
in which $T_{\mu\nu}^{\left(  b\right)  }=\rho_{b}u_{\mu}u_{\nu}$ describes
baryonic matter, $T_{\mu\nu}^{\left(  r\right)  }=\left(  \rho_{r}%
+p_{r}\right)  u_{\mu}u_{\nu}$ corresponds to  radiation,~$T_{\mu\nu
}^{\left(  m\right)  }=\rho_{m}u_{\mu}u_{\nu}$ describes  DM 
component and $T_{\mu\nu}^{\left(  d\right)  }=\left(  \rho_{d}+p_{d}\right)
u_{\mu}u_{\nu}+p_{d}g_{\mu\nu}$ is the energy-momentum tensor for dark
energy fluid source which has a negative equation of state parameter
$w_{d}= p_{d}/\rho_{d}<0$ such that the late-time
accelerating phase of the Universe is realized \cite{ar1}. Here  $u^{\mu}$ describes a
comoving observer $u^{\mu}=\delta_{t}^{\mu},~u^{\mu}u_{\mu}=-1$ and $\rho_i$ ($i= b, r, m, d$) stands for the energy density of $i$-th fluid; $p_r$ and $p_d$ denote the pressure of  radiation and DE fluid. We note that baryons and DM are assumed to be pressureless. 

For spatially flat FLRW background space the gravitational field equations
(\ref{gp.02}) read %
\begin{equation}
3H^{2}=\rho_{m}+\rho_{d}+\rho_{r}+\rho_{b}, \label{gp.04}%
\end{equation}
\begin{equation}
-2\dot{H}-3H^{2}=p_{d}+p_{r}, \label{gp.05}%
\end{equation}
where an overhead dot represents the derivative with respect to the cosmic time and $H=\dot{a}/a$ \ is the Hubble function. From  Bianchi identity $\nabla_{\nu}G^{\mu\nu}=0$, we get  $\nabla_{\nu}T^{\mu\nu}=0,$ which means, 
\begin{equation}
\nabla_{\nu}T^{\left(  m\right)  \mu\nu}+\nabla_{\nu}T^{\left(  d\right)
\mu\nu}+\nabla_{\nu}T^{\left(  r\right)  \mu\nu}+\nabla_{\nu}T^{\left(
b\right)  \mu\nu}=0. \label{gp.06}%
\end{equation}
Often, it is assumed that the fluid components do not interact with each other, that means, 
$\nabla_{\nu}T^{\left(  m\right)  \mu\nu}=0$~,~$\nabla_{\nu
}T^{\left(  d\right)  \mu\nu}=0$~,~$\nabla_{\nu}T^{\left(  r\right)  \mu\nu
}=0$ and $\nabla_{\nu}T^{\left(  b\right)  \mu\nu}=0$. However, over the last several years, cosmological 
models allowing an interaction between DM and DE have drawn a significant attention. In this case the conservation equation (\ref{gp.06}) gives the 
following conditions $~\nabla_{\nu}T^{\left(  r\right)  \mu\nu
}=0$ ,$~\nabla_{\nu}T^{\left(  b\right)  \mu\nu}=0$ and~$\nabla_{\nu
}T^{\left(  m\right)  \mu\nu}+\nabla_{\nu}T^{\left(  d\right)  \mu\nu}=0$ where the last relation between DM and DE  can be decoupled into two separated equations by introducing an interaction term $Q$ as follows%
\begin{equation}
\nabla_{\nu}T^{\left(  m\right)  \mu\nu}=Q~,~\nabla_{\nu}T^{\left(  d\right)
\mu\nu}=-Q. \label{gp.07}%
\end{equation}
We note that the nature of interaction function $Q$ is unknown and there is not a
unique theoretical model which describes its origin
\cite{int01,int02,int03,int04}. In terms of phenomenology, there are various
approaches in the literature which have shown that a nonzero function $Q$ is
supported by the cosmological observations
\cite{int1,int2,int3,int4,int5,int6,int7,int8}.

For the background space (\ref{gp.01}), conservation equations (\ref{gp.07}) take the forms%
\begin{align}
\dot{\rho}_{m}+3H\rho_{m}  &  = Q,\label{gp.08}\\
\dot{\rho}_{d}+3H\left(  \rho_{d}+p_{d}\right)   &  =-Q. \label{gp.09}%
\end{align}
Thus, once an interaction function $Q$ is given, one can solve the conservation equations either analytically or numerically (depending on the nature of the interaction function) and finally using the Friedmann equation (\ref{gp.04}), the dynamics of interacting Universe can be explored. Let us make an important comment that is essential to understand the dynamics in the following sections. In the total energy density of the Universe (see eqn. (\ref{gp.04})) the contributions of  baryonic matter and radiation fluid are very small compared to that of DM and DE, therefore, the total dynamics of the Universe is not influenced by their joint contribution. Hence, we shall omit their contribution from now on and  we shall focus on the dynamics of the Universe driven mainly by  DM and 
DE. That means, the effective cosmological fluid responsible for the Universe's dynamics is identified with the energy-momentum tensor $T_{\mu\nu}^{\rm eff}=T_{\mu\nu}^{\left(  m\right)  }+T_{\mu\nu}^{\left(  d\right)  }$.

As mentioned earlier, the DE fluid is  quintessence scalar field,  one of the simplest time-varying dark energy
models \cite{ra1}. In quintessence DE,   energy density and   pressure terms
of$~T_{\mu\nu}^{\left(  d\right)  }$ are written as
\begin{equation}
\rho_{d}=\frac{1}{2}\dot{\phi}^{2}+V\left(  \phi\right)  ~,~p_{d}=\frac{1}%
{2}\dot{\phi}^{2}-V\left(  \phi\right),  \label{gp.11}%
\end{equation}
where $V (\phi)$ is the potential of quintessence scalar field. With the above expressions for energy density and pressure of the quintessence DE, the set of equations (\ref{gp.08}), (\ref{gp.09}) can now be explicitly written as%
\begin{align}
\dot{\rho}_{m}+3H\rho_{m}  &  =Q,\label{gp.12}\\
\dot{\phi}\left(  \ddot{\phi}+3H\dot{\phi}+V_{,\phi}\right)   &  =-Q.
\label{gp.13}%
\end{align}
The latter system has been the case of study of various analysis in the
literature \cite{am0,am1,am2,am3,am4,am5,am6,am7,am8,am9,am10,am11} where
$Q=Q\left(  \rho_{m},\rho_{d}\right)  $ has been assumed to be either a linear
function $Q=\alpha H\rho_{m}+~\kappa~H\rho_{d}~$ or a nonlinear function such
as  $Q=\alpha\frac{H\rho_{m}}{\dot{\phi}^{2}\rho_{\phi}}~,~Q=\alpha
\frac{H\dot{\phi}^{2}\rho_{\phi}}{\rho_{m}}~$,  $Q=-\frac{1}{2}(4-3 \gamma) \rho_m \dot{\phi}\frac{\chi'(\phi)}{\chi(\phi)}~$ \cite{Leon:2008de,Leon:2010ai,Leon:2014bta,Fadragas:2014mra,int06,int07} and many others. Additionally, with the above choices for the coupling function, various
functional forms of the scalar field potential function $V\left(  \phi\right)
~$ can also be chosen \cite{Gonzalez:2006cj,Leon:2009dt,Cid:2015pja,Cid:2017wtf,int05,am11,dd1}.

In this work we are interested to extend the analysis of  IDE models for a generalization of the quintessence scalar field model inspired by GUP. 

\section{Quintessence modified by the GUP}
\label{sec3}

In this section we offer a brief description about GUP and the modifications of  quintessence scalar field due to this principle. 
The quadratic GUP is based on the introduction of a minimum measurable length 
in which the Heisenberg uncertainty principle is modified as
\begin{equation}
\Delta X_{i}\Delta P_{j}\geqslant\frac{\hbar}{2}[\delta_{ij}(1+\beta
P^{2})+2\beta P_{i}P_{j}]\,, \label{gp.14}%
\end{equation}
where $\beta=\beta_{0}\ell_{Pl}^{2}/2\hbar^{2}~$and $\beta_{0}~$is the
deformation parameter \cite{Quesne2006,Vagenas,Kemph1,Kemph2}. The deformation
parameter can be positive or negative \cite{neg1,neg2}. Moreover, there are various
proposed theoretical constraints in the literature from quantum systems
\cite{neg3,neg4} and also from gravitational systems \cite{neg5,neg6}.

The main characteristic of GUP is that it modifies the second-order Klein-Gordon
equation for a scalar field 
into a fourth-order. Inspired by this
observation in Ref. \cite{gup1}   a generalized
quintessence scalar field model modified by the GUP has been proposed; in which Klein-Gordon equation is modified.

The Action Integral for the scalar field has been proposed to be \cite{gup1}
\begin{equation}
S_{Q}^{GUP}=\int dx^{4}\sqrt{-g}\left(  \frac{1}{2}g^{\mu\nu}\mathcal{D}_{\mu
}\phi\mathcal{D}_{\nu}\phi-V(\phi)\right)  , \label{gp.15}%
\end{equation}
in which$~\mathcal{D}_{\mu}=\nabla_{\mu}+\beta\hbar^{2}\nabla_{\mu}\left(
\Delta\right)  ~$and $\Delta$ is the Laplace operator. For zero deformation
parameter Action Integral (\ref{gp.15}) reduces to that of quintessence. For
$\beta_{0}\neq0$, with the use of a Lagrange multiplier the Action Integral
(\ref{gp.15}) is written in the equivalent form
\begin{equation}
S_{Q}^{GUP}=\int dx^{4}\sqrt{-g}\left(  \frac{1}{2}g^{\mu\nu}\nabla_{\mu}%
\phi\nabla_{\nu}\phi-V(\phi)+\beta\hbar^{2}\left(  2g^{\mu\nu}\nabla_{\mu}%
\phi\nabla_{\nu}\psi+\psi^{2}\right)  \right)  \,, \label{gp.16}%
\end{equation}
where the new scalar field $\psi$ attributes the degrees of freedom provided by higher-order derivatives.

Therefore, in this context DE components read
\begin{equation}
\rho_{d}=\rho_{\phi}+\rho_{\psi}~,~p_{d}=p_{\phi}+p_{\psi}\,, \label{gp.17}%
\end{equation}
where~$\rho_{\phi},~p_{\phi}$ are   usual terms of energy density and
pressure for quintessence scalar field given by (\ref{gp.11}). The quantities $\rho_{\psi}$, $p_{\psi}$ are respectively the energy density and pressure terms which
correspond to  higher-order derivatives and are described by  kinematic and
dynamic quantities for the second field $\psi$ as follows%
\begin{equation}
\rho_{\psi}=\beta\hbar^{2}\left(  2\dot{\phi}\dot{\psi}-\psi^{2}\right), ~p_{\psi}=\beta\hbar^{2}\left(  2\dot{\phi}\dot{\psi}+\psi^{2}\right)  .
\label{gp.18}%
\end{equation}
Hence, using (\ref{gp.17}) and (\ref{gp.18}) the interacting
conservation equations (\ref{gp.12}), (\ref{gp.13}) are modified as
\begin{align}
\dot{\rho}_{m}+3H\rho_{m}  &  =Q,\label{gp.19}\\
2\beta\hbar^{2}\dot{\phi}\left(  \ddot{\psi}+3H\dot{\psi}\right)  +\dot{\phi
}\left(  \psi+V_{,\phi}\right)   &  =-Q\,,\label{gp.20}%
\end{align}
with constraint equation%
\begin{equation}
\ddot{\phi}+3H\dot{\phi}-\psi=0, \label{gp.21}%
\end{equation}
which follows from definition of Lagrange multiplier, where the dot means derivative with respect to cosmic time $t$. 

We continue our analyses by studying the dynamics and  asymptotic behaviour of the solutions of the gravitational field equations for various interacting functions of the form 
$Q=Q\left(  \rho_{m},\rho_{d}\right)  $ in the light of GUP. In particular, we classify stationary points according to their  stability and determine generic features irrespective of the initial conditions. This is achieved using dynamical systems tools. Concerning the choice of  interaction function we consider its linear form:  
\begin{eqnarray}
Q_{A}
=\frac{\dot{\phi}}{\sqrt{6}}\left(  \alpha\rho_{m}+\kappa~\rho_{d}\right), \label{model-Q_A}
\end{eqnarray}
where $\alpha$ and $\kappa$ are  coupling  parameters of the interaction function. Notice that for different values of  $\alpha$ and $\kappa$, one can realize a number of interaction functions. For instance, $\alpha =0$ reduces to $Q_{A}
=\frac{\dot{\phi}}{\sqrt{6}}\; \kappa \rho_{d}$. Similarly, one can also get another two versions, namely, $Q_{A}
=\frac{\dot{\phi}}{\sqrt{6}}\; \alpha\rho_{m}$ for $\kappa =0$ and $Q_{A}
=\frac{\dot{\phi}}{\sqrt{6}} \delta \left( \rho_{m}+\rho_{d}\right)$ for $\alpha = \kappa = \delta$. These kind of interactions were motivated from scalar-tensor theories \cite{Wetterich:1994bg,Amendola:1999qq}. In particular, model  $Q_{A}
=\frac{\dot{\phi}}{\sqrt{6}}\; \alpha\rho_{m}$ is conformally equivalent to  power-law potential model of Brans-Dicke  theory.
As far as the scalar field potential is concerned we consider the exponential
potential $V\left(  \phi\right)  =V_{0}e^{-\lambda\phi}$, which as it was found in \cite{gup2} can describe different kinds of asymptotic solutions with physical interest.

\section{Asymptotic solutions and stability}

\label{sec4}

To study asymptotic evolution of interacting gravitational
model  we define dimensionless variables  following \cite{gup2}:
\begin{equation}
x_{1}=\frac{\dot{\phi}}{\sqrt{6}H}~,~y_{1}=\sqrt{\frac{V}{3H^{2}}}%
~,~x_{2}=\beta\hbar^{2}\frac{2\sqrt{2}\dot{\psi}}{\sqrt{3}H}~,~y_{2}%
=\frac{\beta\hbar^{2}\psi^{2}}{3H^{2}},~\Omega_{m}=\frac{\rho_{m}}{3H^{2}}. 
\label{gp.22}%
\end{equation}
Using the new variables, the field equations (\ref{gp.04}),
(\ref{gp.05}), (\ref{gp.12}), (\ref{gp.13}) reduce to an
algebraic-differential system of first-order. We determine the stationary points of the new system  and investigate their stability. Every stationary point describes an asymptotic solution for the cosmological field equations.
\subsection{Interaction $Q_{A}$}
For interaction model $Q_{A}$, the field equations reduce to the equivalent
system%
\begin{eqnarray}
\frac{dx_{1}}{d\tau}&= &\frac{1}{4}\left(  6x_{1}\left( x_1(x_1+x_2)-y_1^{2}-1\right)  +y_2\left(  6x_1-\sqrt{6}\mu\right)  \right)  , \label{gp.23}\\
\frac{dy_{1}}{d\tau} &= &\frac{1}{2}\,y_1\left( x_1\,\left(
-\sqrt{6}\lambda+3\,x_1+3x_2\right)  -3y_1^{2}+3y_2+3\right)  , \label{gp.24}\\
\frac{d x_2}{d\tau} & =& \frac{1}{2}\left(-2\alpha\left(  -x_{1}\left(  x_1+x_2\,\right)  -y_{1}^{2}+y_{2}+1\right)  +x_1(3x_2-2\kappa)\left( x_1+x_2\right)\right. \nonumber\\
&&  \left. +y_{1}^{2}\left(  2\sqrt{6}\lambda-2\kappa-3 x_2 \right)+y_{2}\left(  \sqrt{6}\mu+2\kappa + 3x_2\right)  -3 x_2 \right)\,,\label{gp.25}\\
\frac{dy_{2}}{d\tau}&=&\frac{1}{4}y_2 \left(  12 x_1^{2}+12\, x_{1}x_{2}-\sqrt{6}\mu\,x_2+12\left(  -y_1^{2}+y_2+1\right)  \right)  , \label{gp.26}\\
\frac{d\mu}{d\tau}&=&\frac{1}{4}\sqrt{\frac{3}{2}}\mu^{2}\,x_2\,, \label{gp.27}
\end{eqnarray}
with the constraint equation%
\begin{equation}
\Omega_{m}=1-\left(  x_{1}^{2}+y_{1}^{2}\right)  -\left(  x_{1}x_{2}%
-y_{2}\right)  , \label{gp.28}%
\end{equation}
where the new phase space variable $\mu$ is defined as $\mu=-2\left(  \beta\hbar^{2}%
\psi\right)  ^{-1}$. 
From the requirement that the dimensionless energy densities of matter fluids must be  non-negative, i.e., $\Omega_m\geq 0$ and $\Omega_d:= \left(  x_{1}^{2}+y_{1}^{2}\right)  +\left(  x_{1}x_{2}-y_{2}\right)\geq 0$ we have the physical part of the phase space is $0 \leq \left(  x_{1}^{2}+y_{1}^{2}\right)  +\left(  x_{1}x_{2}%
-y_{2}\right)\leq 1$. 
However, it is important to mention here, that from expression (\ref{gp.28}) the phase space variables are not bounded, but the set of initial conditions in which parameters $x_2, y_2$ reach infinity are not physically acceptable (because $x_2, y_2$ are proportional to $\beta$ and $\beta\ll 1$ from the model's construction). For this reason one can omit the analysis when $x_2$ or $y_2$ tends to infinity.

In the new variables the effective equation of state parameter $w_{\rm eff}%
=-1-\frac{2}{3}\frac{\dot{H}}{H^{2}}$ is written as
\begin{equation}
w_{\rm eff}\left(  x_{1},x_{2},y_{1},y_{2}\right)  =x_{1}^{2}-y_{1}^{2}+x_{1}%
x_{2}+y_{2}. \label{gp.29}%
\end{equation}
Note that at a stationary point the scale factor is expressed as $a\left(
t\right)  =a_{0}t^{\frac{2}{3\left(  1+w_{\rm eff}\right)  }},$ for $w_{\rm eff}%
\neq-1$ and $a\left(  t\right)  =a_{0}e^{H_{0}t}$ for $w_{\rm eff}=-1$.
On the other hand, according to \eqref{gp.18}, we have 
\begin{equation}
w_{\psi}= \frac{p_{\psi}}{\rho_{\psi}}= \frac{\left(  2\dot{\phi}\dot{\psi}-\psi^{2}\right)}{\left(  2\dot{\phi}\dot{\psi}+\psi^{2}\right)}= \frac{x_1 x_2-y_2}{x_1 x_2+y_2}.
\label{gp.29}%
\end{equation}
The deceleration parameter  $q%
=-1-\frac{\dot{H}}{H^{2}}$ is written as 
\begin{equation}
    q= \frac{1}{2} (3 w_{\rm eff}+1)=  \frac{1}{2} \left(3 x_1 (x_1+x_2)-3 y_1^2+3
   y_2+1\right). 
\end{equation}
It is useful to define  $c_{d}^2$, the effective sound speed of DE
perturbations (the corresponding quantity for matter is
zero in the dust case) as  
\begin{align}
  &  c_{d}^2 = w_{d} - \frac{w_{d}'}{3(1+w_{d})}\,,
  \end{align}
  where $w_d=\frac{p_d}{\rho_d}$ and prime denotes derivative with respect to $\tau$.  In terms of phase-space variables we have 
  \begin{align}
 & c_{d}^2=   \frac{x_1 x_2^2 \left(12 x_1-\sqrt{6} \mu  y_2\right)+x_2 y_2 \left(12 x_1+\sqrt{6} \mu  y_2\right)-2 \sqrt{6} x_1 y_2 \left(2 \lambda  y_1^2+\mu 
   y_2\right)}{12 x_1 x_2 (x_1 x_2+y_2)}\nonumber \\
   & -\frac{\alpha  y_2 \left(x_1 (x_1+x_2)+y_1^2-y_2-1\right)}{3 x_2 (x_1 x_2+y_2)}+\frac{\kappa
    y_2 \left(x_1 (x_1+x_2)+y_1^2-y_2\right)}{3 x_2 (x_1 x_2+y_2)}.
\end{align}

Before studying the general case for arbitrary values of the coupling parameters $\left\{  \alpha,\kappa\right\}  $ in the following sections we consider the special cases where $\kappa=0$ or $\alpha=0$.

\subsubsection{Interaction $Q_{A}$ with $\kappa=0$}

For $\kappa=0$, the dynamical system consisted by the differential equations
(\ref{gp.23})-(\ref{gp.27}) admits the stationary points $\mathbf{A}%
=\mathbf{A}\left(  \mathbf{x,y,}\mu;\lambda,\alpha\right)  $, $\mathbf{x=}%
\left(  x_{1},x_{2}\right)  ,~\mathbf{y}=\left(  y_{1},y_{2}\right)  $ with
coordinates%
\begin{align*}
A_{0}^{\pm}  &  =\left(\pm 1,0,0,0,\mu\right) ~,~A_{1}    =\left(  x_{1},\frac{1}{x_{1}}-x_{1},0,0,0\right)  ~,~A_{2}%
=\left(  0,\sqrt{\frac{2}{3}}\lambda y_{1}^{2},y_{1},y_{1}^{2}-1,0\right)  ,\\
A_{3}  &  =\left(  0,-\frac{2}{3}\alpha,0,0,0\right)  ~,~A_{4}=\left(
\frac{3}{\alpha},0,0,-1-\frac{9}{\alpha^{2}},\frac{6\sqrt{6}\alpha}%
{9+\alpha^{2}}\right)  .
\end{align*}
While the family of points $A_0^{\pm}$ constitutes a line,  $A_{1},~A_{2}$ describe surfaces in the phase space. We continue with
the presentation of  physical properties of the stationary points and investigate asymptotic solutions' stability properties. To determine the stability behavior, we examine the nature of   eigenvalues corresponding to a perturbed matrix of the system \eqref{gp.23}-\eqref{gp.27} for each point.

The family of points $A_{0}^{\pm}$ describe asymptotic solutions where only the
kinetic parts of  scalar field contribute to the effective cosmological
fluid, that is, $\Omega_{m}\left(  A_{0}^\pm\right)  =0$ and $w_{\rm eff}\left(
A_{0}^\pm \right)  =1$. Hence, the asymptotic solution for the scale factor is
$a\left(  t\right)  =a_{0}t^{\frac{1}{3}}$. The eigenvalues of the linearized
system around  each stationary point are $e_{1}\left(  A_{0}^\pm\right)
=6$,~$e_{2}\left(  A_{0}^\pm\right)  =3\pm \alpha $, $e_{3}\left(  A_{0}^\pm\right)
=\frac{1}{2}\left(  6\mp \sqrt{6}\lambda \right)  $, $e_{4}\left(A_{0}^\pm\right)  =0$ and $e_{5}\left(  A_{0}^\pm\right)  =0$. Therefore, the family of points $A_{0}^{+}$ is  unstable when $ \alpha\,>-3$ and $ \lambda\,<\sqrt{6}$, otherwise it is saddle. Also, the family of points $A_{0}^{-}$ is  unstable when $ \alpha\,<3$ and $ \lambda\,>-\sqrt{6}$, otherwise it is saddle.

The family of points $A_{1}$ describe  the same physical properties as that of $A_0^\pm$. The family of  points $A_1$ and $A_{0}^\pm$ intersect at  points $(\pm 1,0,0,0,0)$. The eigenvalues of the linearized
system around the stationary point are $e_{1}\left(  A_{1}\right)
=6$,~$e_{2}\left(  A_{1}\right)  =3+\alpha x_{1}$, $e_{3}\left(  A_{1}\right)
=\frac{1}{2}\left(  6-\sqrt{6}\lambda x_{1}\right)  $, $e_{4}\left(
A_{1}\right)  =0$ and $e_{5}\left(  A_{1}\right)  =0$. Looking  from the nature of eigenvalues,  the family of points $A_{1}$ is  unstable when $\alpha\,x_1>-3$ and $\lambda\,x_1<\sqrt{6}$, otherwise it is saddle as one of
the eigenvalues is always positive. Therefore, the asymptotic solutions at $A_{1}$ are  not stable.

The family of points $A_{2}$ describes  de Sitter asymptotic solution where the dynamical part of scalar field
dominates in the field equations, that is, $\Omega_{m}\left(  A_{2}\right)
=0,~w_{\rm eff}\left(  A_{2}\right)  =-1$ and $a\left(  t\right)  =a_{0}e^{H_{0}%
t}$. The eigenvalues of the linearized system are $e_{1}\left(
A_{2}\right)  =-3$, $e_{2}\left(  A_{2}\right)  =-3$,~$e_{3}\left(
A_{2}\right)  =-3$, $e_{4}\left(  A_{2}\right)  =0$, $e_{5}\left(
A_{2}\right)  =0$.  Therefore, we apply the center manifold theorem (CMT) to analyze the stability of $A_2$.

Point $A_{3}$ describes a
universe dominated by DM fluid $\Omega_{m}\left(  A_{3}\right)
=1~w_{\rm eff}\left(  A_{3}\right)  =0$ with scale factor $a\left(  t\right)
=a_{0}t^{\frac{2}{3}}$. The eigenvalues of the linearized system are
$e_{1}\left(  A_{3}\right)  =3$, $e_{2}\left(  A_{3}\right)  =-\frac{3}{2}%
$,~$e_{3}\left(  A_{3}\right)  =-\frac{3}{2}$, $e_{4}\left(  A_{3}\right)
=\frac{3}{2}$, $e_{5}\left(  A_{3}\right)  =0$. Therefore,  the asymptotic solution is  not stable and the point is a  saddle. 

Finally, for point $A_{4}$ we find $\Omega_{m}\left(  A_{4}\right)  =-\frac{18}{\alpha^{2}}<0$ and so the point is not physically acceptable. Therefore it is not necessary to investigate its stability.

Compared with the corresponding model of usual quintessence theory \cite{Boehmer:2008av}, the present model does not contain any accelerated scaling solution. However, the present model exhibits new stationary points describing late time de Sitter solution and matter-dominated solution.

\begin{figure}[ptb]
\centering\includegraphics[width=1\textwidth]{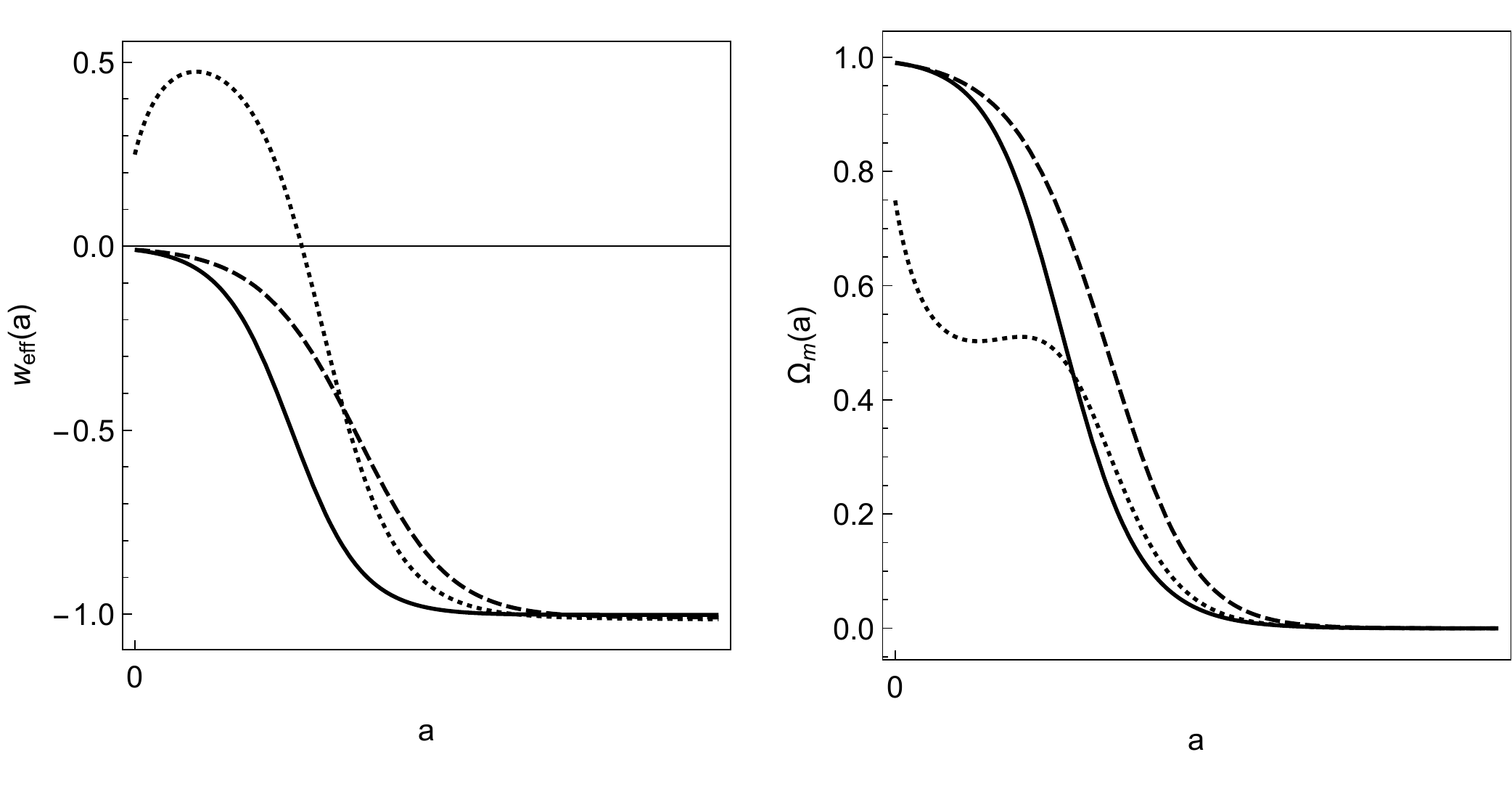} \caption{Qualitative
evolution of  $w_{\rm eff}\left(  a\right)  $ (left figure) and $\Omega
_{m}\left(  a\right)  $ (right figure) for the interacting model  $Q_A$ with
$\kappa=0$. Solid line $\left(  \alpha=5\right)  $ and dashed line $\left(
\alpha=-5\right)  $ are for the initial conditions $\left(  x_{1}^{0},x_{2}%
^{0},y_{1}^{0},y_{2}^{0},\mu^{0}\right)  =\left(  0,-\frac{2}{3}%
\alpha,0,-0.01.0.2\right)  $,~while the dotted line is for $\alpha=-5$ with initial conditions $\left(  x_{1}^{0},x_{2}^{0},y_{1}^{0},y_{2}^{0},\mu
^{0}\right)  =\left(  0.5,0,0,0.01,0.2\right)  $. The unique attractor of the
dynamical system is point $A_{2}$ which describes the de Sitter universe.}%
\label{fig1A}%
\end{figure}

In Fig. \ref{fig1A} we present the qualitative evolution of the physical
parameters $w_{\rm eff}\left(  a\right)  $ and $\Omega_{m}\left(  a\right)  $ for
various sets of initial conditions near to the matter dominated era. Note that we have chosen initial conditions where  the final attractor is the de Sitter solution described by the point $A_{2}$.

\subsubsection{Interaction $Q_{A}$ with $\alpha=0$}

For arbitrary parameter $\kappa$ and $\alpha=0$, the stationary points of the
dynamical system (\ref{gp.23})-(\ref{gp.27}) are 
\begin{align*}
\bar{A}_{1}  &  =\left(  0,0,0,0,\mu\right)  ~,~\bar{A}_{2}=\left(
0,\frac{\sqrt{6}\lambda y_{1}^{2}-\kappa}{3},y_{1},y_{1}^{2}-1,0\right)  ,\\
\bar{A}_{3}  &  =\left(  \frac{\sqrt{6}}{\lambda},\frac{\left(  \lambda
^{2}-12\right)  \kappa-\sqrt{6}\lambda\left(  \lambda^{2}-6\right)  }%
{2\lambda\left(  \sqrt{6}\kappa-3\lambda\right)  },\sqrt{\frac{\kappa\left(
9\lambda-2\sqrt{6}\kappa\right)  }{2\sqrt{6}\kappa^{2}-30\kappa\lambda
+9\sqrt{6}\lambda^{2}}},0,0\right)  ,\\
\bar{A}_{4}^{\left(  \pm\right)  }  &  =\left(  -\frac{3\pm\sqrt{9-2\kappa
^{2}}}{2\kappa},0,0,-\frac{9+\kappa^{2}\pm3\sqrt{9-2\kappa^{2}}}{2\kappa^{2}%
},-\frac{2\sqrt{6}\kappa\left(  9\pm\sqrt{9-2\kappa^{2}}\right)  }%
{36+\kappa^{2}}\right)  .
\end{align*}
The family of stationary points $\bar{A}_{1}$ describes a universe dominated by dark
matter fluid, that is, $\Omega_{m}\left(  \bar{A}_{1}\right)  =1,~w_{\rm eff}%
\left(  \bar{A}_{1}\right)  =0$ and $a\left(  t\right)  =a_{0}t^{\frac{2}{3}}%
$. The eigenvalues of the linearized system are $e_{1}\left(  \bar{A}%
_{1}\right)  =3$, $e_{2}\left(  \bar{A}_{1}\right)  =-\frac{3}{2}$%
,~$e_{3}\left(  \bar{A}_{1}\right)  =-\frac{3}{2}$, $e_{4}\left(  \bar{A}%
_{1}\right)  =\frac{3}{2}$, $e_{5}\left(  \bar{A}_{1}\right)  =0$ which means
that the point is a saddle point and the asymptotic solution is not stable.

$\bar{A}_{2}$ describes a family of stationary points of de Sitter asymptotic
solutions, with physical properties $\Omega_{m}\left(  \bar{A}_{2}\right)
=0,~w_{\rm eff}\left(  \bar{A}_{2}\right)  =-1$ and $a\left(  t\right)
=a_{0}e^{H_{0}t}$. The eigenvalues are exactly that of $A_{2}$, therefore we shall apply CMT.

For point $\bar{A}_{3}$, we find $\Omega_{m}\left(  \bar{A}_{3}\right)  =\frac{2\kappa}{2\kappa-\sqrt{6}\lambda
},~w_{\rm eff}\left(  \bar{A}_{3}\right)  =1$. Therefore, for the point to be real and physically acceptable $\left\{  y_{1}\left(
\bar{A}_{3}\right)  \geq0,~0\leq \Omega_{m}\left(  \bar{A}_{3}\right)  \leq1\right\}  $,  we must have $\kappa=0$ i.e. there is no interaction. Hence, we omit the study of its stability.

Finally, for the point $\bar{A}_{4}^{\left(  \pm\right)  }$, we obtain $\Omega
_{m}\left(  \bar{A}_{4}^{\left(  \pm\right)  }\right)  =1-\frac{9}{\kappa^{2}%
}\mp\frac{3}{\kappa^{2}}\sqrt{9-2\kappa^{2}}$, $w_{\rm eff}\left(  \bar{A}%
_{4}^{\left(  \pm\right)  }\right)  =-1$. However, the point is not physically acceptable as there is no value of $\kappa$ satisfying the physical conditions $\left\{0\leq\Omega_{m}\left(  \bar{A}_{4}^{\left(  \pm\right)  }\right)  \leq 1,9-2\kappa^{2}\geq 0\,\right\} $.

\subsubsection{Interaction $Q_{A}$ with arbitrary $\alpha$ and $\kappa$}
In the general case where the coupling parameters $\alpha$~and $\kappa$ are
arbitrary we find the stationary points
\begin{align*}
A_{1}^{\prime}  &  =\left(  0,-\frac{2\alpha}{3},0,0,0\right)  ~,~A_{2}%
^{\prime}=\left(  0,\frac{\sqrt{6}\lambda y_{1}^{2}-\kappa}{3},y_{1},y_{1}%
^{2}-1,0\right)  ~,\\
A_{3}^{\prime}  &  =\left(  \frac{\sqrt{6}}{\lambda},\frac{\left(  \lambda
^{2}-6\right)  \left(  2\alpha+\sqrt{6}\lambda\right)  -\left(  \lambda
^{2}-12\right)  \kappa}{2\lambda\left(  \sqrt{6}\left(  \alpha-\kappa\right)
+3\lambda\right)  },\sqrt{\frac{\kappa\left(  2\sqrt{6}\left(  \alpha
-\kappa\right)  +9\lambda\right)  }{4\sqrt{6}\left(  \alpha-\kappa\right)
^{2}+3\lambda\left(  10\left(  \alpha-\kappa\right)  +3\sqrt{6}\lambda\right)
}},0,0\right)  ,\\
A_{4}^{\prime\left(  \pm\right)  }  & =\left(  \frac{3\pm K_{4}}{2\left(
\alpha-\kappa\right)  },0,0,-\frac{9+2\alpha^{2}-3\alpha\kappa+\kappa^{2}%
\pm3K_{4}}{2\left(  \alpha-\kappa\right)  ^{2}},-\frac {2\sqrt {6} \left(  \left( \kappa-2\,\alpha \right)  \left( 3\pm K_4
 \right) +6\,\kappa \right) }{ \left( 2\,\alpha-\kappa
 \right) ^{2}+36}\right)\,.
\end{align*}
where $K_{4}=\sqrt{9+2\kappa\left(  \alpha-\kappa\right)  }$.

The stationary point $A_{1}^{\prime}$ describes a matter dominated solution with $\Omega_{m}\left(
A_{1}^{\prime}\right)  =1,~w_{\rm eff}\left(  A_{1}^{\prime}\right)  =0$ and
asymptotic solution $a\left(  t\right)  =a_{0}t^{\frac{2}{3}}$. The linearized system around the stationary point admits  eigenvalues
$e_{1}\left(  A_{1}^{\prime}\right)  =3$, $e_{2}\left(  A_{1}^{\prime}\right)
=-\frac{3}{2}$,~$e_{3}\left(  A_{1}^{\prime}\right)  =-\frac{3}{2}$,
$e_{4}\left(  A_{1}^{\prime}\right)  =\frac{3}{2}$, $e_{5}\left(
A_{1}^{\prime}\right)  =0$ from where we infer that $A_{1}^{\prime}$ is a
saddle point. 

The family of stationary points described by $A_{2}^{\prime}$ have the same
physical properties and eigenvalues as that of family of points $A_{2}$ and $\bar{A}_{2}$. Hence, we shall apply the CMT  to investigate the stability of the points. 

Point $A_{3}^{^{\prime}}$ describes an asymptotic solution with
$\Omega_{m}\left(  A_{3}^{\prime}\right)=-\frac{2\kappa}{\sqrt{6} \lambda+2(\alpha-\kappa)}$ and $w_{\rm eff}\left(  A_{3}^{\prime}\right)  =1$. We can easily find that the asymptotic solution at the stationary point is physically
accepted only when $\kappa=0$, which belongs  to a family of points $A_{1}$  for $x_1=\frac{\sqrt{6}}{\lambda}$.

For a stationary point $A_{4}^{\prime\left(  \pm\right)  }$, we find that $\Omega_{m}\left(  A_{4}^{\prime\left(  \pm\right)  }\right)  =\frac{\kappa^{2}-a\kappa-3\left(  3\pm K_{4}\right)  }{\left(  \alpha
-\kappa\right)  ^{2}}$,~$w_{\rm eff}\left(  A_{4}^{\prime\left(  \pm\right)
}\right)  =-1$. However, this point is not physically acceptable as there are no real values for $\kappa$ and $\alpha$ such that $0\leq \Omega_m\leq 1$. In the next subsection, we shall proceed with the analysis of a non-hyperbolic equilibrium point $A_{2}^{^{\prime}}$.\\

\subsubsection{CMT for $A_{2}^{\prime}$}

We use the Center Manifold Theorem to analyze the stability of the stationary point $A_{2}^{^{\prime}}$ of system  (\ref{gp.23})-(\ref{gp.27}) and we discuss the specific
cases $\kappa=0,~$or $\alpha=0$ which correspond to the points $A_{2}$ and
$\bar{A}_{3}$ respectively.

Firstly, we assume $(\sqrt{6} \kappa + 6 (1 - 2 y_c^2) \lambda)\neq 0, y_c\neq0, y_c^2\neq 1$. Then, evaluating the Jacobian matrix of system (\ref{gp.23})-(\ref{gp.27}) at $A_{2}^{^{\prime}}$ 
we obtain 
\begingroup\makeatletter\def\f@size{6.5}\check@mathfonts
\begin{align}
  J(A_{2}^{^{\prime}})= \scriptscriptstyle \left(
\begin{array}{ccccc}
 -3 & 0 & 0 & 0 & -\frac{1}{2} \sqrt{\frac{3}{2}} \left(y_1^2-1\right) \\
 \frac{1}{2} y_1 \left(\sqrt{6} \lambda  \left(y_1^2-1\right)-\kappa \right) & -3 y_1^2 & 0 & \frac{3 y_1}{2} & 0 \\
 -\frac{1}{6} \left(\kappa -\sqrt{6} \lambda  y_1^2\right) \left(2 \alpha -3 \kappa +\sqrt{6} \lambda  y_1^2\right) & -y_1 \left(-2 \alpha +\kappa +\sqrt{6} \lambda  \left(y_1^2-2\right)\right) & -3 &
   \frac{1}{2} \left(-2 \alpha +\kappa +\sqrt{6} \lambda  y_1^2\right) & \sqrt{\frac{3}{2}} \left(y_1^2-1\right) \\
 \left(y_1^2-1\right) \left(\sqrt{6} \lambda  y_1^2-\kappa \right) & -6 y_1 \left(y_1^2-1\right) & 0 & 3 \left(y_1^2-1\right) & \frac{1}{12} \left(y_1^2-1\right) \left(\sqrt{6} \kappa -6
   \lambda  y_1^2\right) \\
 0 & 0 & 0 & 0 & 0 \\
\end{array}
\right)\,,
\end{align}
\endgroup
which can be reduced to its canonical Jordan Form
\begin{align}
j(A_{2}^{^{\prime}})=   \left(
\begin{array}{ccccc}
 -3 & 1 & 0 & 0 & 0 \\
 0 & -3 & 1 & 0 & 0 \\
 0 & 0 & -3 & 0 & 0 \\
 0 & 0 & 0 & 0 & 1 \\
 0 & 0 & 0 & 0 & 0 \\
\end{array}
\right)\,,
\end{align}
defining a similarity matrix $s$ which satisfies $J(A_{2}^{^{\prime}})= s j(A_{2}^{^{\prime}}) s^{-1}$. 
The eigenvalues of $J(A_{2}^{^{\prime}})$ and $j(A_{2}^{^{\prime}})$ are  $-3,-3,-3,0,0$.  Defining new variables
\begin{align}
 ( v_1, v_2, v_3, u_1, u_2)=  s^{-1}.\left(x_1,y_1-y_c,x_2+\frac{1}{3} \left(\kappa -\sqrt{6} \lambda  y_c^2\right),y_2-y_c^2+1,\mu \right),
\end{align}
which translates a fixed point on the curve $A_{2}^{^{\prime}}$ labelled by the value of $y_c$ to the origin, that is,
\begin{subequations}
\label{aux}
\begin{align}
  & v_1=\frac{1}{216} \Bigg(12 \Big(3 \sqrt{6} \mu +y_c^2 \Big(-\left(3 \sqrt{6} \mu +\lambda  \left(\lambda  \left(y_c^2-1\right) \left(12 x_1+\sqrt{6} \mu  \left(3 y_c^2-2\right)\right)+6 \sqrt{6}
   \left(y_2+y_c^2\right)\right)\right)\Big) \nonumber \\
   & +12 \sqrt{6} \lambda  y_1 \left(y_c^2-1\right) y_c\Big)   +\kappa  \left(72-2 \mu  \left(y_c^2-1\right) \left(\sqrt{6} \alpha +3 \lambda 
   y_c^2\right)\right)+3 \sqrt{6} \kappa ^2 \mu  \left(y_c^2-1\right)\Bigg)+x_2,
\\
  & v_2= \frac{x_1 \left(\kappa -\sqrt{6} \lambda  \left(y_c^2-1\right)\right)  \left(2 \alpha -\kappa +\sqrt{6} \lambda  y_c^2\right)^2}{6 \left(\kappa  (\kappa -2 \alpha )+\sqrt{6} \lambda  \left(\kappa +2 \alpha 
   \left(y_c^2-1\right)-2 \kappa  y_c^2\right)+6 \lambda ^2 y_c^2 \left(y_c^2-1\right)\right)^2}  \nonumber \\
   &\times \Bigg(\kappa ^2 (3 \kappa -2 \alpha )+6 \lambda
   ^2 y_c^2 \left(\kappa  \left(3 y_c^2-2\right)-2 \alpha  \left(y_c^2-1\right)\right) \nonumber \\
   & +\sqrt{6} \kappa  \lambda  \left(\alpha  \left(4 y_c^2-2\right)+\kappa  \left(3-7 y_c^2\right)\right)+6 \sqrt{6}
   \lambda ^3 y_c^2 \left(y_c^4-3 y_c^2+2\right)\Bigg) \nonumber \\
   & +y_c (y_1-y_c) \left(2 \alpha -\kappa +\sqrt{6} \lambda  y_c^2\right)+\frac{1}{2}
   \left(y_2-y_c^2+1\right) \left(-2 \alpha +\kappa -\sqrt{6} \lambda  y_c^2\right) \nonumber \\
   & -\frac{\mu  \left(y_c^2-1\right)  \left(2 \alpha -\kappa +\sqrt{6}
   \lambda  y_c^2\right)}{72 \left(\kappa  (\kappa -2 \alpha )+\sqrt{6} \lambda  \left(\kappa +2 \alpha  \left(y_c^2-1\right)-2 \kappa  y_c^2\right)+6 \lambda ^2 y_c^2
   \left(y_c^2-1\right)\right)^2} \nonumber \\
   & \times \Bigg(3 \sqrt{6} \kappa ^5-24 \kappa ^2 \lambda  \left(y_c^2-1\right) \left(2
   \alpha ^2+\sqrt{6} \alpha  \lambda  \left(5 y_c^2-2\right)+3 \lambda ^2 y_c^2 \left(3 y_c^2-1\right)\right) \nonumber \\
   & +12 \sqrt{6} \kappa  \lambda ^2 \left(y_c^2-1\right)^2 \left(2 \alpha ^2+3 \lambda ^2
   y_c^2 \left(2-3 y_c^2\right)\right)-4 \kappa ^4 \left(2 \sqrt{6} \alpha +9 \lambda  \left(2 y_c^2-1\right)\right) \nonumber \\
   & +144 \lambda ^4 y_c^2 \left(y_c^2-1\right)^3 \left(\sqrt{6} \alpha +3 \lambda 
   y_c^2\right)+2 \kappa ^3 \left(2 \sqrt{6} \alpha ^2+24 \alpha  \lambda  \left(3 y_c^2-2\right)+3 \sqrt{6} \lambda ^2 \left(16 y_c^4-16 y_c^2+3\right)\right)\Bigg),
\\
   & v_3= \frac{1}{24} \kappa  \left(12 x_1+\sqrt{6} \mu  \left(y_c^2-1\right)\right) \left(-2 \alpha +\kappa -\sqrt{6} \lambda  y_c^2\right),
\\
   & u_1= \frac{1}{3} y_c \left(y_c
   \left(\left(y_c^2-1\right) \left(\sqrt{6} \lambda  x_1+\lambda  \mu  \left(y_c^2-1\right)+3\right)+3 y_2\right)-6 y_1 \left(y_c^2-1\right)\right),\\
   & u_2=\frac{1}{12} \mu  y_c^2
   \left(y_c^2-1\right) \left(\sqrt{6} \kappa +6 \lambda  \left(1-2 y_c^2\right)\right)\,, \end{align}
\end{subequations}
we obtain evolution  equations $u_1', u_2', v_1', v_2'$ and $v_3'$ from the equations that results from taking derivative with respect to $\tau$ of \eqref{aux} and substituting $x_1', y_1', x_2', y_2'$ and $\mu'$ from equations (\ref{gp.23})-(\ref{gp.27}) and taking inverse transformation of \eqref{aux}. The linear part of the resulting system is written as
    \begin{align}
\left(
\begin{array}{ccccc}
 v_1' \\
 v_2' \\
 v_3' \\
 u_1' \\
 u_2' \\
\end{array}
\right)=   \left(
\begin{array}{ccccc}
 -3 & 1 & 0 & 0 & 0 \\
 0 & -3 & 1 & 0 & 0 \\
 0 & 0 & -3 & 0 & 0 \\
 0 & 0 & 0 & 0 & 1 \\
 0 & 0 & 0 & 0 & 0 \\
\end{array}
\right)\left(
\begin{array}{ccccc}
 v_1 \\
 v_2 \\
 v_3 \\
 u_1 \\
 u_2 \\
\end{array}
\right). 
\end{align}
Therefore, the center manifold is given by the graph 
\begin{align}
\label{XgraphmanifoldD}
 W^{c}_{loc}(\mathbf{0})=  &  \Bigg\{(v_1, v_2, v_3,u_1, u_2)\in \mathbb{R}^5: v_i=h_i(u_1, u_2), \nonumber\\
 & h_i(0,0)=0, 
  \frac{\partial h_i}{\partial u_j} \Bigg{|}_{(u_1,u_2)=(0,0)}= 0, i=1,2,3, j=1,2, ||(u_1, u_2)||<\delta\Bigg\} \,,
\end{align}
for some $\delta>0$. 
\newline 
The $h_i$'s satisfy a set of partial quasi-linear differential equations than can be written symbolically by 
\begin{subequations}
\label{pdes}
\begin{align}
 & -u_2' \frac{\partial h_1}{\partial u_2}(u_1,u_2)-u_1' \frac{\partial h_1}{\partial u_1}(u_1,u_2)+v_1'=0,\\
  & -u_2' \frac{\partial h_2}{\partial u_2}(u_1,u_2)-u_1'
   \frac{\partial h_2}{\partial u_1}(u_1,u_2)+v_2',\\
   & -u_2' \frac{\partial h_3}{\partial u_2}(u_1,u_2)-u_1' \frac{\partial h_3}{\partial u_1}(u_1,u_2)+v_3'=0.
\end{align}
\end{subequations}
where the prime means derivative with respect to $\tau$. These equations are
obtained by substituting the expressions $u_1', u_2', v_1', v_2'$ and $v_3'$ from the equations that results from taking derivative with respect to $\tau$ of \eqref{aux} and substituting $x_1', y_1', x_2', y_2'$ and $\mu'$ from equations (\ref{gp.23})-(\ref{gp.27})  and taking inverse transformation of \eqref{aux}. Next, replacing $v_1\mapsto h_1(u_1, u_2), v_2\mapsto h_2(u_1, u_2), v_3\mapsto h_3(u_1, u_2)$ we obtain the final equations for $h_1(u_1, u_2), \; h_2(u_1, u_2), \; h_3(u_1, u_2)$. 
In general, we solve these equations by using Taylor expansions in the independent variables $u_1, u_2$ that are of order greater or equal than two. Therefore, we propose the expansion in series 
\begin{equation}
\label{taylor}
    h_i(u_1, u_2)= \sum_{n=2}^{N}\sum_{k=0}^{n} a^{[i]}_{n k} u_1^{n-k} u_2^{k} + \mathcal{O}(\|(u_1,u_2)^{N+1}\|),\;\; i=1, 2, 3.  
\end{equation}
\begin{figure}[ptb]
\centering\includegraphics[width=0.8\textwidth]{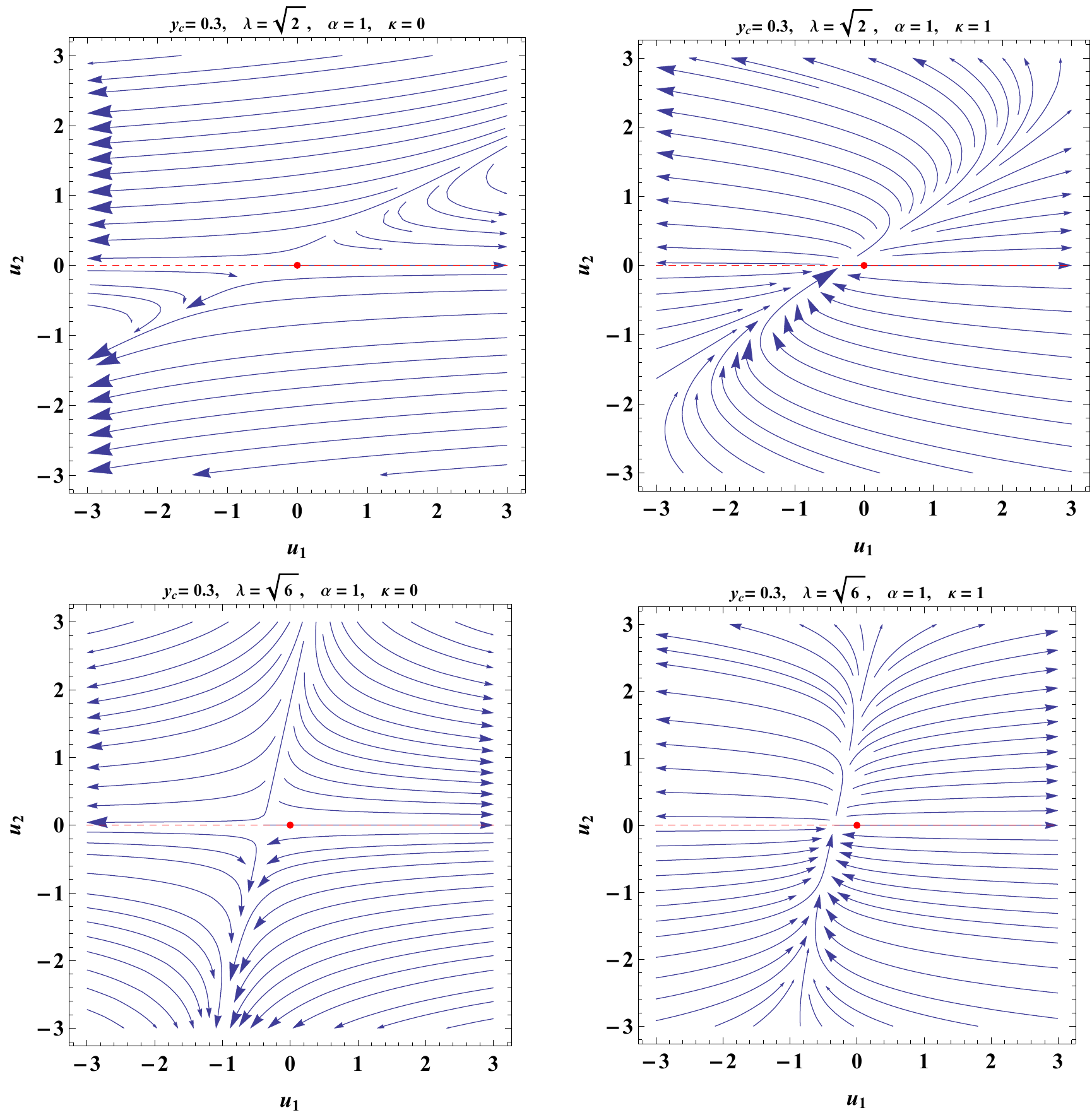}
    \caption{Phase plot of system \eqref{dscenter} for some values of the parameters showing that $A_{2}^{^{\prime}}$ is a saddle point.}
    \label{fig:my_label}
\end{figure}
Setting $N=2$, we obtain the coefficients 
$a^{[1]}_{20}= -\frac{\left(y_c^2-1\right) \lambda }{2 \sqrt{6} y_c^2}, \; a^{[2]}_{20}= \frac{-\sqrt{6} \lambda  y_c^2-2 \alpha +\kappa }{8 y_c^2}, \; a^{[3]}_{20}= 0, \; a^{[3]}_{21}= -\frac{\kappa  \left(-6 \lambda  y_c^2-2 \sqrt{6} \alpha +\sqrt{6} \kappa \right)}{2 y_c^2 \left(y_c^2-1\right) \left(\sqrt{6} \kappa
   +6 \left(1-2 y_c^2\right) \lambda \right)},$ and 
   $a^{[1]}_{21}, a^{[2]}_{21}, a^{[1]}_{22}, a^{[2]}_{22}$ and $a^{[3]}_{22}$ are complicated expressions of the parameters. 
   \newline 
Finally, the dynamics on the center manifold up to third order is given by a system of the form 
\begin{subequations}
\label{dscenter}
\begin{align}
& u_1'=  \alpha_1  {u_2^2}+ \alpha_2  {u_1  u_2}+u_2  + \mathcal{O}(\|(u_1,u_2)^{3}\|),
   \\
 &  u_2'= -\frac{u_2^2 \left(\sqrt{6} \kappa -6 y_c^2 \lambda \right)}{2 y_c^2 \left(y_c^2-1\right) \left(\sqrt{6} \kappa +6 \left(1-2 y_c^2\right) \lambda \right)} + \mathcal{O}(\|(u_1,u_2)^{3}\|) \,,\label{centerb}
\end{align}
\end{subequations}
where $\alpha_1$ and $\alpha_2$ are complicated functions of the parameters.

From equation \eqref{centerb} the center is unstable  to perturbations along the $u_2$-axis (saddle behavior). This is due to \eqref{centerb} which is a gradient like equation 
\begin{equation}
    u_2'=-U'(u_2), \quad U(u_2)= \frac{u_2^3 \left(\sqrt{6} \kappa -6 y_c^2 \lambda \right)}{6 y_c^2 \left(y_c^2-1\right) \left(\sqrt{6} \kappa +6 \left(1-2 y_c^2\right) \lambda \right)},
\end{equation}
such that $u_2=0$ is an inflection point of $U(u_2)$. That is, the solution either depart or approach the origin along the $u_2$-axis, depending on the sign of the initial value of $u_2$,  $u_{20}$ and the sign of $\frac{\left(\sqrt{6} \kappa -6 y_c^2 \lambda \right)}{  y_c^2 \left(y_c^2-1\right) \left(\sqrt{6} \kappa +6 \left(1-2 y_c^2\right) \lambda \right)}$. The system also admits an invariant line of stationary points which at second order is approximated by $\alpha_1 u_2 + \alpha_2 u_1 +1=0$ which behaves as saddle.

We represent the phase plot of the system \eqref{dscenter} for several choices of parameters in Fig.  \ref{fig:my_label} where the origin is a saddle point. 

Now, we study the particular cases $(\sqrt{6} \kappa + 6 (1 - 2 y_c^2) \lambda)= 0$, or  $y_c=0$,  or $y_c^2=1$.

\paragraph{Subcase $\lambda=\frac{\kappa }{\sqrt{6} \left(2 y_c^2-1\right)}$, $y_c^2\neq \frac{1}{2}$.} In this subcase the similarity matrix is 
\begingroup\makeatletter\def\f@size{6}\check@mathfonts
\begin{align}
    s= \left(
\begin{array}{ccccc}
 -\frac{y_c^2-1}{2 \sqrt{6}} & 0 & 0 & 0 & \frac{1}{72 \kappa  \left(2 y_c^2-1\right)^3 \left(\alpha  \left(2-4 y_c^2\right)+\kappa  \left(y_c^2-1\right)\right)} \\
 \frac{\kappa  y_c \left(y_c^2-1\right)}{12 \sqrt{6} \left(2 y_c^2-1\right)} & \frac{1}{2 y_c} & 0 & \frac{y_c}{12 \left(2 y_c^2-1\right) \left(\kappa +\alpha  \left(4
   y_c^2-2\right)-\kappa  y_c^2\right)} & \frac{\kappa  y_c \left(y_c^2-1\right)^2}{216 \left(1-2 y_c^2\right)^4 \left(\kappa +\alpha  \left(4 y_c^2-2\right)-\kappa  y_c^2\right)^2} \\
 \frac{\left(y_c^2-1\right) \left(2 \alpha  \kappa +(y_c-1) (y_c+1) \left(3 \kappa ^2+8 y_c^2 (\kappa  (\alpha -\kappa )+18)\right)+36\right)}{36 \sqrt{6} \left(1-2 y_c^2\right)^2} & \frac{\kappa
   }{6 y_c^2-3} & 1 & 0 & 0 \\
 0 & 1 & 0 & \frac{y_c^2-1}{6 \left(2 y_c^2-1\right) \left(\kappa +\alpha  \left(4 y_c^2-2\right)-\kappa  y_c^2\right)} & \frac{\left(y_c^2-1\right) \left(\alpha  \left(4 y_c^2-2\right)+\kappa
    \left(2 y_c^4-5 y_c^2+3\right)\right)}{216 \left(1-2 y_c^2\right)^4 \left(\kappa +\alpha  \left(4 y_c^2-2\right)-\kappa  y_c^2\right)^2} \\
 1 & 0 & 0 & 0 & 0 \\
\end{array}
\right)\,,
\end{align}
\endgroup
and the real Jordan form of the matrix is 
\begin{align}
 j(A_{2}^{^{\prime}})=   \left(
\begin{array}{ccccc}
 0 & 0 & 0 & 0 & 0 \\
 0 & 0 & 0 & 0 & 0 \\
 0 & 0 & -3 & 1 & 0 \\
 0 & 0 & 0 & -3 & 1 \\
 0 & 0 & 0 & 0 & -3 \\
\end{array}
\right).
\end{align}
Defining 
\begin{align*}
(u_1, u_2, v_1, v_2, v_3 )= s^{-1} \cdot    \left(x_1,y_1-y_c,x_2+\frac{1}{3} \left(\kappa -\frac{\kappa  y_c^2}{2 y_c^2-1}\right),y_2-y_c^2+1,\mu \right) ,
\end{align*}
and using the CMT we obtain that the center manifold is given by \eqref{XgraphmanifoldD}, where the $h_i$ satisfy \eqref{pdes}. \newline 
Using the ansatz \eqref{taylor} with $N=2$, we obtain the coefficients \newline
$a^{[1]}_{20}= -\frac{\kappa  \left(y_c^2-1\right)^2 \left(-4 \alpha ^2+12 \alpha  \kappa -9 \kappa ^2+8 y_c^6 \left(4 \alpha ^2-11 \alpha  \kappa +7 \kappa ^2-72\right)+y_c^4 \left(-48 \alpha ^2+136 \alpha 
   \kappa -101 \kappa ^2+720\right)+y_c^2 \left(24 \alpha ^2-70 \alpha  \kappa +54 \kappa ^2-288\right)+36\right)}{2592 \left(2 y_c^2-1\right)^3}$, \newline
   $a^{[2]}_{20}= \frac{\left(y_c^2-1\right)^2 \left(144
   y_c^4+\left(\kappa ^2-144\right) y_c^2+36\right) \left(\kappa +\alpha  \left(4 y_c^2-2\right)-\kappa  y_c^2\right)}{144 \left(2 y_c^2-1\right)}, a^{[3]}_{20}= 0, a^{[1]}_{21}= \frac{2 \alpha  \kappa -\kappa
   ^2+y_c^4 \left(8 \alpha  \kappa +3 \left(\kappa ^2+48\right)\right)-2 y_c^2 \left(4 \alpha  \kappa +\kappa ^2+72\right)+36}{36 \sqrt{6} \left(1-2 y_c^2\right)^2}$,
   \newline
   $a^{[2]}_{21}= \frac{\kappa  \left(\alpha  \left(20
   y_c^4-22 y_c^2+6\right)+\kappa  \left(-9 y_c^4+16 y_c^2-7\right)\right)}{2 \sqrt{6}}, a^{[3]}_{21} =6 \sqrt{6} \kappa  \left(2 y_c^2-1\right)^3 \left(\kappa +\alpha  \left(4 y_c^2-2\right)-\kappa
    y_c^2\right)$,
   \newline
   $a^{[1]}_{22}= -\frac{\kappa -\kappa  y_c^2}{12 y_c^2-24 y_c^4}, a^{[2]}_{22}= -\frac{3 \left(2 y_c^2-1\right) \left(\kappa +\alpha  \left(4 y_c^2-2\right)-\kappa  y_c^2\right)}{2
   y_c^2}, a^{[3]}_{22}= 0$.

\begin{figure}[ptb]
\centering\includegraphics[width=0.8\textwidth]{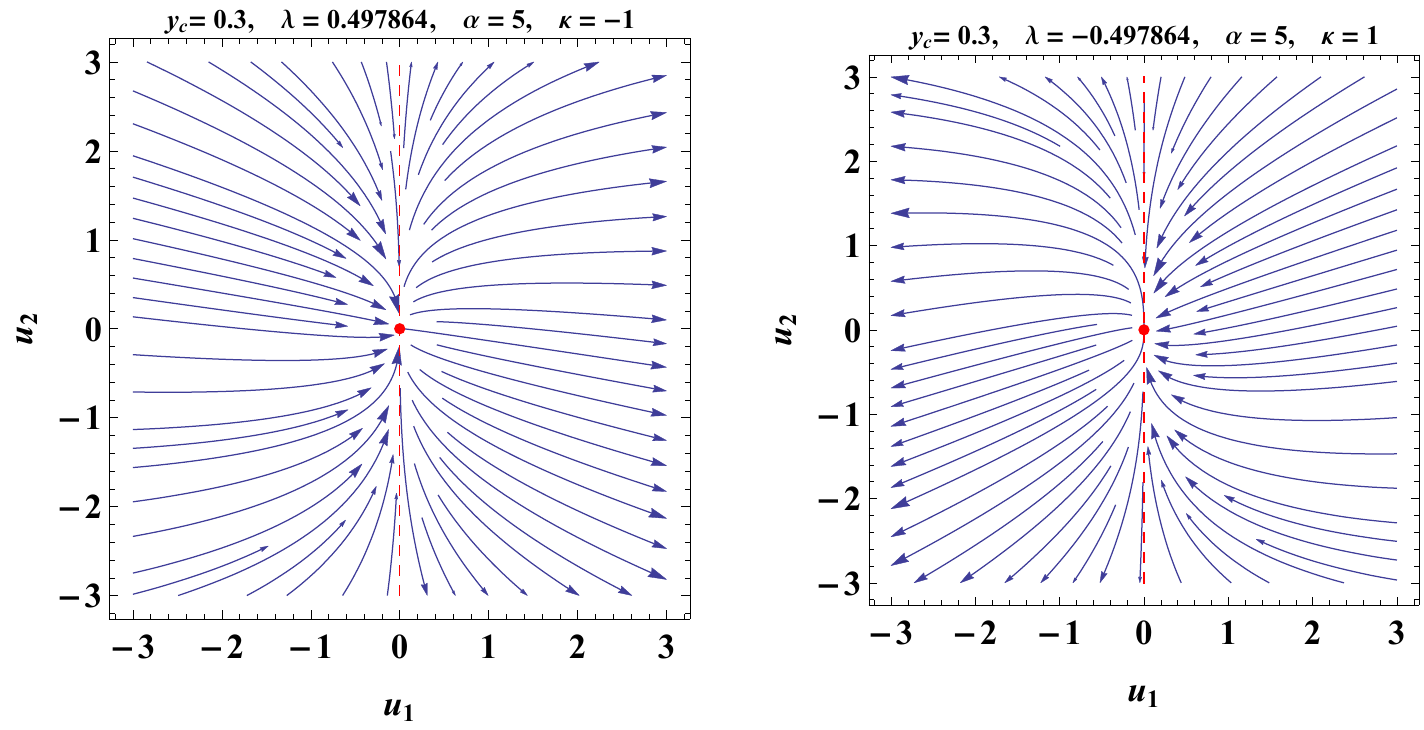}
    \caption{Phase plot of system \eqref{dscenter2} for some values of the parameters showing that $A_{2}^{^{\prime}}$ is a saddle point.}
    \label{fig:my_label2}
\end{figure}

Therefore, the dynamics on the center manifold is governed up to third order by 
\begin{subequations}
\label{dscenter2} 
\begin{align}
 & u_1'=  \frac{\sqrt{6} \kappa  u_1^2 \left(y_c^2-1\right)}{24(1-2 y_c^2)}, \label{centerc}\\
 & u_2'=-\frac{u_1^2 \left(y_c^2-1\right)^2 y_c^2 \left(-\alpha  \kappa +2 \left(\kappa ^2-9\right)+2 y_c^2 (\kappa 
   (\alpha -\kappa )+18)\right)}{72 \left(2 y_c^2-1\right)}-\frac{\kappa  u_1 u_2 \left(y_c^2-1\right) y_c^2}{\sqrt{6} \left(2 y_c^2-1\right)}.
\end{align}
\end{subequations}
From equation \eqref{centerc} the center is unstable  to perturbations along the $u_1$-axis (saddle behavior). This is due to \eqref{centerc} which is a gradient like equation 
\begin{equation}
    u_1'=-U'(u_1), \quad U(u_1)=- \frac{\sqrt{6} \kappa  u_1^3 \left(y_c^2-1\right)}{72(1-2 y_c^2)},
\end{equation}
such that $u_1=0$ is an inflection point of $U(u_1)$. That is, the solution either depart or approach the origin along the $u_1$-axis, depending on the sign of the initial value of $u_1$,  $u_{10}$ and the sign of $\frac{\kappa  \left(y_c^2-1\right)}{(1-2 y_c^2)}$.

In Fig. \ref{fig:my_label2} a phase plot of system \eqref{dscenter2} is presented for some values of the parameters showing that $A_{2}^{^{\prime}}$ is a saddle point.

\paragraph{Subcase $\kappa=0, y_c^2= \frac{1}{2}$.} 

For $y_c= \frac{\sqrt{2}}{2}$  the similarity matrix is
\begin{align}
  s=\left(
\begin{array}{ccccc}
 \frac{1}{\frac{\lambda ^2}{4 \alpha +\sqrt{6} \lambda }+\frac{\lambda }{\sqrt{6}}} & 0 & \frac{1}{\sqrt{\frac{2}{3}} \alpha  \lambda +\lambda ^2} & \frac{1}{4 \sqrt{6}} & 0 \\
 -\frac{3 \lambda }{4 \sqrt{3} \alpha +6 \sqrt{2} \lambda } & 0 & \frac{1}{2 \sqrt{2} \alpha +2 \sqrt{3} \lambda } & -\frac{\lambda }{24 \sqrt{2}} & \frac{1}{\sqrt{2}} \\
 0 & 1 & 0 & -\frac{\lambda ^2+12}{24 \sqrt{6}} & \sqrt{\frac{2}{3}} \lambda  \\
 1 & 0 & 0 & 0 & 1 \\
 0 & 0 & 0 & 1 & 0 \\
\end{array}
\right) \,, 
\end{align}
and the real Jordan matrix is 
\begin{align}
\label{matrix50}
 j(A_{2}^{^{\prime}})=  \left(
\begin{array}{ccccc}
 -3 & 0 & 0 & 0 & 0 \\
 0 & -3 & 1 & 0 & 0 \\
 0 & 0 & -3 & 0 & 0 \\
 0 & 0 & 0 & 0 & 0 \\
 0 & 0 & 0 & 0 & 0 \\
\end{array}
\right).
\end{align}
In this case we define 
\begin{subequations}
\begin{align*}
& v_1=\frac{1}{24} \left(-\lambda  \mu +2 \sqrt{6} \lambda  x_1-12 \sqrt{2} y_1+12 y_2+18\right), \\
& v_2=  \frac{1}{144} \left(-\sqrt{6} \left(\lambda ^2-12\right) \mu -12 \lambda  \left(-2 \lambda  x_1+4 \sqrt{3}
   y_1+2 \sqrt{6} y_2+\sqrt{6}\right)\right)+x_2,\\
& v_3=\alpha  \left(\frac{\lambda  x_1}{\sqrt{6}}+\sqrt{2} y_1-y_2-\frac{3}{2}\right)-\frac{1}{48} \lambda  \left(\sqrt{6} (\lambda  \mu +18)-36
   \lambda  x_1-24 \sqrt{3} y_1+12 \sqrt{6} y_2\right),\\
& u_1= \mu,\\
& u_2= \frac{1}{24} \left(\lambda  \mu -2 \sqrt{6} \lambda  x_1+12 \sqrt{2} y_1+12 y_2-6\right).
\end{align*}
\end{subequations}
On the other hand, for $y_c=- \frac{\sqrt{2}}{2}$ the similarity matrix changes to \begin{align}
  s=\left(
\begin{array}{ccccc}
 \frac{1}{\frac{\lambda ^2}{4 \alpha +\sqrt{6} \lambda }+\frac{\lambda }{\sqrt{6}}} & 0 & \frac{1}{\sqrt{\frac{2}{3}} \alpha  \lambda +\lambda ^2} & \frac{1}{4 \sqrt{6}} & 0 \\
 \frac{3 \lambda }{4 \sqrt{3} \alpha +6 \sqrt{2} \lambda } & 0 & \frac{1}{-2 \sqrt{2} \alpha -2 \sqrt{3} \lambda } & \frac{\lambda }{24 \sqrt{2}} & -\frac{1}{\sqrt{2}} \\
 0 & 1 & 0 & -\frac{\lambda ^2+12}{24 \sqrt{6}} & \sqrt{\frac{2}{3}} \lambda  \\
 1 & 0 & 0 & 0 & 1 \\
 0 & 0 & 0 & 1 & 0 \\
\end{array}
\right) \,,
\end{align}
and the real Jordan matrix is \eqref{matrix50}.
In this case we define
\begin{subequations}
\begin{align*}
& v_1=\frac{1}{24} \left(-\lambda  \mu +2 \sqrt{6} \lambda  x_1+12 \sqrt{2} y_1+12 y_2+18\right),\\
& v_2= \frac{1}{144} \left(-\sqrt{6} \left(\lambda ^2-12\right) \mu -12 \lambda  \left(-2 \lambda  x_1-4 \sqrt{3}
   y_1+2 \sqrt{6} y_2+\sqrt{6}\right)\right)+x_2,\\
   & v_3= -\frac{1}{6} \alpha  \left(-\sqrt{6} \lambda  x_1+6 \sqrt{2} y_1+6 y_2+9\right)-\frac{1}{48} \lambda  \left(\sqrt{6} (\lambda  \mu +18)-36
   \lambda  x_1+24 \sqrt{3} y_1+12 \sqrt{6} y_2\right),\\
   & u_1= \mu , \\
   & u_2= \frac{1}{24} \left(\lambda  \mu -2 \sqrt{6} \lambda  x_1-12 \sqrt{2} y_1+12 y_2-6\right).
\end{align*}
\end{subequations}
The real Jordan matrix is the same as before.
In both cases the  CMT gives the same result. That is,  the center manifold is given by \eqref{XgraphmanifoldD}, where the $h_i$ satisfy \eqref{pdes}. \newline 
Using the ansatz \eqref{taylor} with $N=2$, we obtain the coefficients  
$a^{[1]}_{20}= \frac{-\lambda ^2-12}{2304}, a^{[2]}_{20}=\frac{\lambda  \left(\lambda ^2+24\right)}{1152 \sqrt{6}}, a^{[3]}_{20}= \frac{\left(\lambda ^2+12\right) \left(4 \alpha +\sqrt{6} \lambda \right)}{4608}, a^{[1]}_{21}= -\frac{\lambda }{16}, a^{[2]}_{21}= \frac{24-5 \lambda ^2}{24 \sqrt{6}}, a^{[3]}_{21}= -\frac{1}{96} \lambda  \left(4 \alpha +5 \sqrt{6} \lambda \right), a^{[1]}_{22}=\frac{1}{4}
   , a^{[2]}_{22}=\frac{\lambda }{2 \sqrt{6}}, a^{[3]}_{22}=\frac{1}{8}
   \left(-4 \alpha -\sqrt{6} \lambda \right)$.
   \newline 
Therefore, the dynamics on the center manifold is governed up to third order by 
\begin{subequations}
\label{dscenter3b} 
\begin{align}
& u_1'= \frac{\lambda  u_1^2}{8}, \label{center3d} \\
& u_2'=\frac{\lambda  u_1 u_2}{4}-\frac{u_1^2}{32}.
\end{align}
\end{subequations}
From equation \eqref{center3d} the center is unstable  to perturbations along the $u_1$-axis (saddle behavior). This is due to \eqref{center3d} which is a gradient like equation 
\begin{equation}
    u_1'=-U'(u_1), \quad U(u_1)= -\frac{\lambda  u_1^3}{24},
\end{equation}
such that $u_1=0$ is an inflection point of $U(u_1)$. That is, the solution either depart or approach the origin along the $u_1$-axis, depending on the sign of the initial value of $u_1$,  $u_{10}$ and the sign of $\lambda$.

\begin{figure}[ptb]
\centering\includegraphics[width=0.8\textwidth]{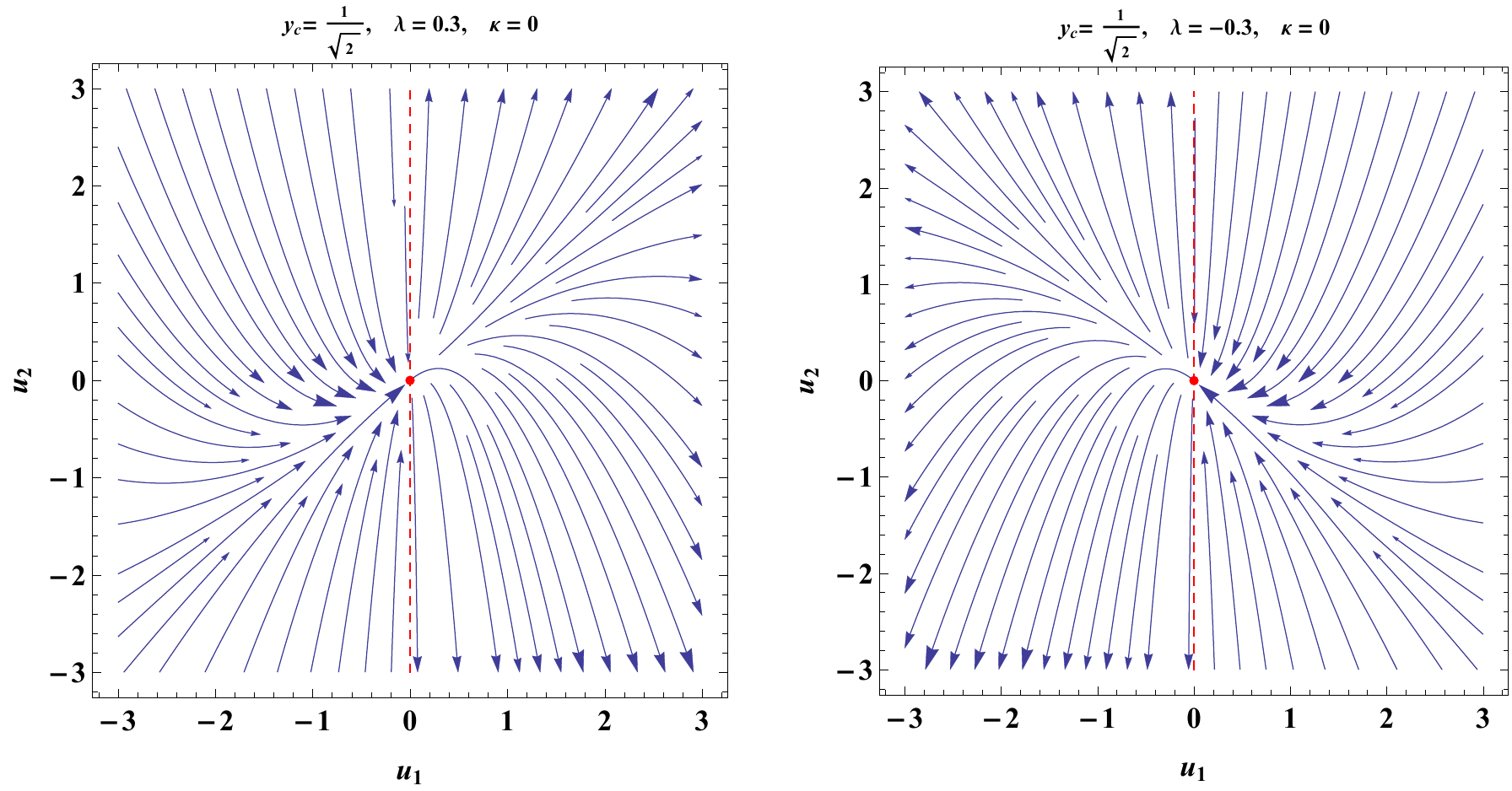}
    \caption{Phase plot of system \eqref{dscenter3b} for some values of the parameters showing that $A_{2}^{^{\prime}}$ is a saddle point.}
    \label{fig:my_label5}
\end{figure}
In Fig.  \ref{fig:my_label5} a phase plot of system \eqref{dscenter3b} is presented for some values of the parameters showing that $A_{2}^{^{\prime}}$ is a saddle point.

\paragraph{Subcase $y_c=0$, $\kappa \neq 2\alpha$.} 
In this subcase the similarity matrix is
\begin{align}
 s=   \left(
\begin{array}{ccccc}
 0 & 0 & -\frac{2}{2 \alpha  \kappa -\kappa ^2} & \frac{1}{2 \sqrt{6}} & 0 \\
 0 & 0 & 0 & 0 & 1 \\
 1 & 0 & 0 & -\frac{2 \alpha  \kappa -3 \kappa ^2+36}{36 \sqrt{6}} & 0 \\
 0 & \frac{2}{\kappa -2 \alpha } & \frac{4 \alpha -6 \kappa }{3 (\kappa -2 \alpha )^2} & 0 & 0 \\
 0 & 0 & 0 & 1 & 0 \\
\end{array}
\right)\,,
\end{align}
and the real Jordan matrix is
\begin{align}
\label{matrix66}
 j(A_{2}^{^{\prime}})=   \left(
\begin{array}{ccccc}
 -3 & 1 & 0 & 0 & 0 \\
 0 & -3 & 1 & 0 & 0 \\
 0 & 0 & -3 & 0 & 0 \\
 0 & 0 & 0 & 0 & 0 \\
 0 & 0 & 0 & 0 & 0 \\
\end{array}
\right).
\end{align}
Defining 
\begin{subequations}
\begin{align*}
& v_1=\frac{\mu  \left(2 \alpha  \kappa -3 \kappa ^2+36\right)}{36 \sqrt{6}}+\frac{\kappa }{3}+x_2, \\
& v_2= \frac{1}{72} \left(3 \kappa  \left(-\sqrt{6} \kappa  \mu +12 \kappa  x_1+12 y_2+12\right)-2 \alpha 
   \left(-\sqrt{6} \kappa  \mu +12 \kappa  x_1+36 y_2+36\right)\right),\\
& v_3= -\frac{1}{24} \kappa  (2 \alpha -\kappa ) \left(12 x_1-\sqrt{6} \mu \right), \;  u_1= \mu,  \;  u_2=y_1\;,
\end{align*}
\end{subequations}
and using the CMT we obtain that the center manifold is given by \eqref{XgraphmanifoldD}, where the $h_i$ satisfy \eqref{pdes}. \newline 
Using the ansatz \eqref{taylor} with $N=2$, we obtain the coefficients 
$a^{[1]}_{20}= \frac{\kappa  \left(-4 \alpha ^2+12 \alpha  \kappa -9 \kappa ^2+36\right)}{2592}, a^{[2]}_{20}= \frac{1}{48} (2 \alpha -\kappa ), a^{[3]}_{20}=0, a^{[1]}_{21}=0, a^{[2]}_{21}=0, a^{[3]}_{21}= 0, a^{[1]}_{22}= \sqrt{\frac{2}{3}} \lambda
   , a^{[2]}_{22}=\frac{1}{2} (\kappa -2 \alpha ), a^{[3]}_{22}=0$. 
Therefore, the dynamics on the center manifold is governed up to third order by 
\begin{subequations}
\label{dscenter3} 
\begin{align}
& u_1'= -\frac{\kappa  u_1^2}{4 \sqrt{6}}, \label{centerd} \\
& u_2'= \frac{1}{24} u_1 u_2 \left(-\sqrt{6} \kappa -6 \lambda \right).
\end{align}
\end{subequations}
From equation \eqref{centerd} the center is unstable  to perturbations along the $u_1$-axis (saddle behavior). This is due to \eqref{centerd} which is a gradient like equation 
\begin{equation}
    u_1'=-U'(u_1), \quad U(u_1)=\frac{\kappa  u_1^3}{12 \sqrt{6}},
\end{equation}
such that $u_1=0$ is an inflection point of $U(u_1)$. That is, the solution either depart or approach the origin along the $u_1$-axis, depending on the sign of the initial value of $u_1$,  $u_{10}$ and the sign of $\kappa$.

\begin{figure}[ptb]
\centering\includegraphics[width=0.8\textwidth]{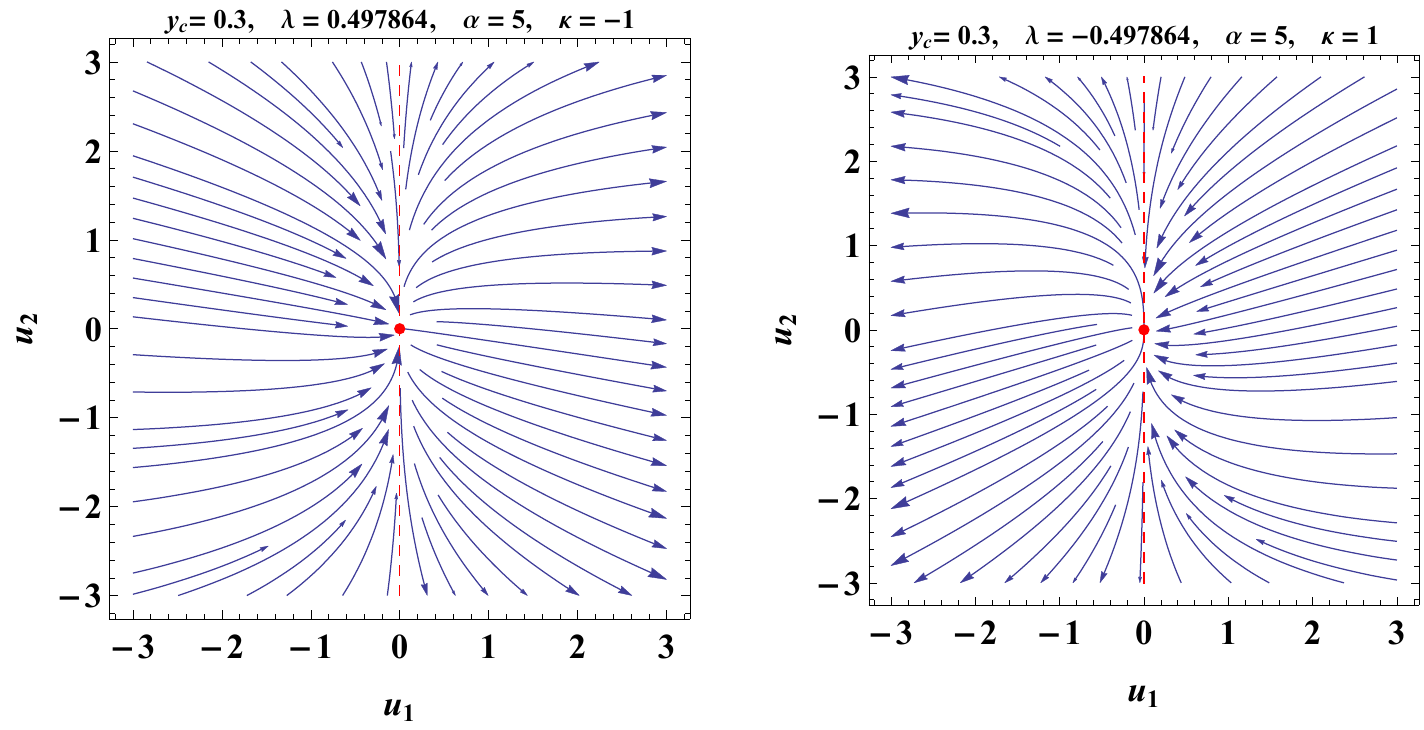}
    \caption{Phase plot of system \eqref{dscenter3} for some values of the parameters showing that $A_{2}^{^{\prime}}$ is a saddle point.}
    \label{fig:my_label3}
\end{figure}
In Fig. \ref{fig:my_label3} a phase plot of system \eqref{dscenter3} is presented for some values of the parameters showing that $A_{2}^{^{\prime}}$ is a saddle point.

\paragraph{Subcase $y_c=0$, $\kappa = 2\alpha$.} 

In this case, the similarity matrix is 
\begin{equation}
 s=   \left(
\begin{array}{ccccc}
 0 & 0 & 1 & \frac{1}{2 \sqrt{6}} & 0 \\
 0 & 0 & 0 & 0 & 1 \\
 -\frac{\kappa }{3} & \frac{\kappa ^2}{3} & 0 & \frac{\kappa ^2-18}{18 \sqrt{6}} & 0 \\
 0 & \kappa  & 0 & 0 & 0 \\
 0 & 0 & 0 & 1 & 0 \\
\end{array}
\right)\,,
\end{equation} and the real Jordan matrix is \eqref{matrix50}.

Defining
\begin{subequations}

\begin{align*}
  & v_1= \frac{\sqrt{6} \left(\kappa ^2-18\right) \mu -108 x_2+36 \kappa  y_2}{36 \kappa }, \;  v_2= \frac{y_2+1}{\kappa }, \;  v_3= x_1-\frac{\mu }{2 \sqrt{6}}, \;  u_1= \mu, \;  u_2= y_1\;,
\end{align*}
\end{subequations}
and using the CMT we obtain that the center manifold is given by \eqref{XgraphmanifoldD}, where the $h_i$ satisfy \eqref{pdes}. \newline 
Using the ansatz \eqref{taylor} with $N=2$, we obtain the coefficients
$a^{[1]}_{20}= \frac{1}{216} \left(\kappa ^2-18\right), a^{[2]}_{20}=  -\frac{1}{24 \kappa }, a^{[3]}_{20}=  0,a^{[1]}_{21}= 0, a^{[2]}_{21}=  0, a^{[3]}_{21}=  0, a^{[1]}_{22}= 1-\frac{\sqrt{6} \lambda }{\kappa }, a^{[2]}_{22}=  \frac{1}{\kappa }, a^{[3]}_{22}=  0$.
Hence, we obtain the system governing the dynamics on the center manifold is \eqref{dscenter3}. Then, it follows the same result as before. That is,  $A_{2}^{^{\prime}}$ is a saddle point.

\paragraph{Subcase $y_c^2=1$, $2 \alpha -\kappa +\sqrt{6} \lambda \neq 0$.} 
In this subcase the similarity matrix is
\begin{align}
 s=   \left(
\begin{array}{ccccc}
 0 & 0 & \frac{2}{\kappa  \left(-2 \alpha +\kappa -\sqrt{6} \lambda \right)} & 0 & 0 \\
 0 & \frac{\epsilon }{2 \alpha -\kappa +\sqrt{6} \lambda } & \frac{\epsilon  \left(\kappa -\sqrt{6} \lambda \right) \left(-2 \alpha +3 \kappa -\sqrt{6} \lambda \right)}{3 \kappa  \left(2 \alpha -\kappa +\sqrt{6} \lambda
   \right)^2} & 0 & \frac{\epsilon }{2} \\
 1 & 0 & 0 & 0 & \sqrt{\frac{2}{3}} \lambda  \\
 0 & 0 & 0 & 0 & 1 \\
 0 & 0 & 0 & 1 & 0 \\
\end{array}
\right)\,,
\end{align}
where $\epsilon=\pm 1$ is the value of $y_c$ and the real Jordan matrix is the same as \eqref{matrix66}.

Defining 
\begin{subequations}
\begin{align*}
& v_1=\frac{1}{3} \left(\kappa +3 x_2-\sqrt{6} \lambda  (y_2+1)\right), \\
& v_2= \frac{1}{6} \left(6 \lambda ^2 x_1-2 \alpha  \left(\kappa  x_1-\sqrt{6} \lambda  x_1-6 y_1 \epsilon +3
   y_2+6\right)-\sqrt{6} \lambda  (4 \kappa  x_1-6 y_1 \epsilon +3 y_2+6)+3 \kappa  (\kappa  x_1-2 y_1 \epsilon +y_2+2)\right) ,\\
& v_3= \frac{1}{2} \kappa  x_1 \left(-2 \alpha +\kappa
   -\sqrt{6} \lambda \right), \;  u_1= \mu, \;  u_2=y_2\,, 
\end{align*}
\end{subequations}
and using the CMT we obtain that the center manifold is given by \eqref{XgraphmanifoldD}, where the $h_i$ satisfy \eqref{pdes}. \newline 
Using the ansatz \eqref{taylor} with $N=2$, we obtain the coefficients  
$a^{[1]}_{20}= 0, a^{[2]}_{20}= 0, a^{[3]}_{20}= 0, a^{[1]}_{21}= \frac{1}{216} \left(2 \sqrt{6} \alpha  \kappa -3 \sqrt{6} \kappa ^2+6 \kappa  \lambda +12 \sqrt{6} \left(\lambda ^2+3\right)\right), a^{[2]}_{21}=\frac{1}{72} \kappa 
   \left(2 \sqrt{6} \alpha -3 \sqrt{6} \kappa +18 \lambda \right), a^{[3]}_{21}= \frac{1}{24} \kappa  \left(2 \sqrt{6} \alpha -\sqrt{6} \kappa +6 \lambda \right), a^{[1]}_{22}=0,a^{[2]}_{22}= \frac{1}{8} \left(-2 \alpha +\kappa -\sqrt{6}
   \lambda \right), a^{[3]}_{22}= 0$. 
Therefore, the dynamics on the center manifold is governed up to third order by 
\begin{subequations}
\label{dscenter4} 
\begin{align}
& u_1'= -\frac{1}{24} u_1^2 \left(\sqrt{6} \kappa -6 \lambda \right), \label{centere} \\
& u_2'= \frac{1}{12} u_1 u_2 \left(\sqrt{6} \kappa -6 \lambda \right).
\end{align}
\end{subequations}
\begin{figure}[ptb]
\centering\includegraphics[width=0.8\textwidth]{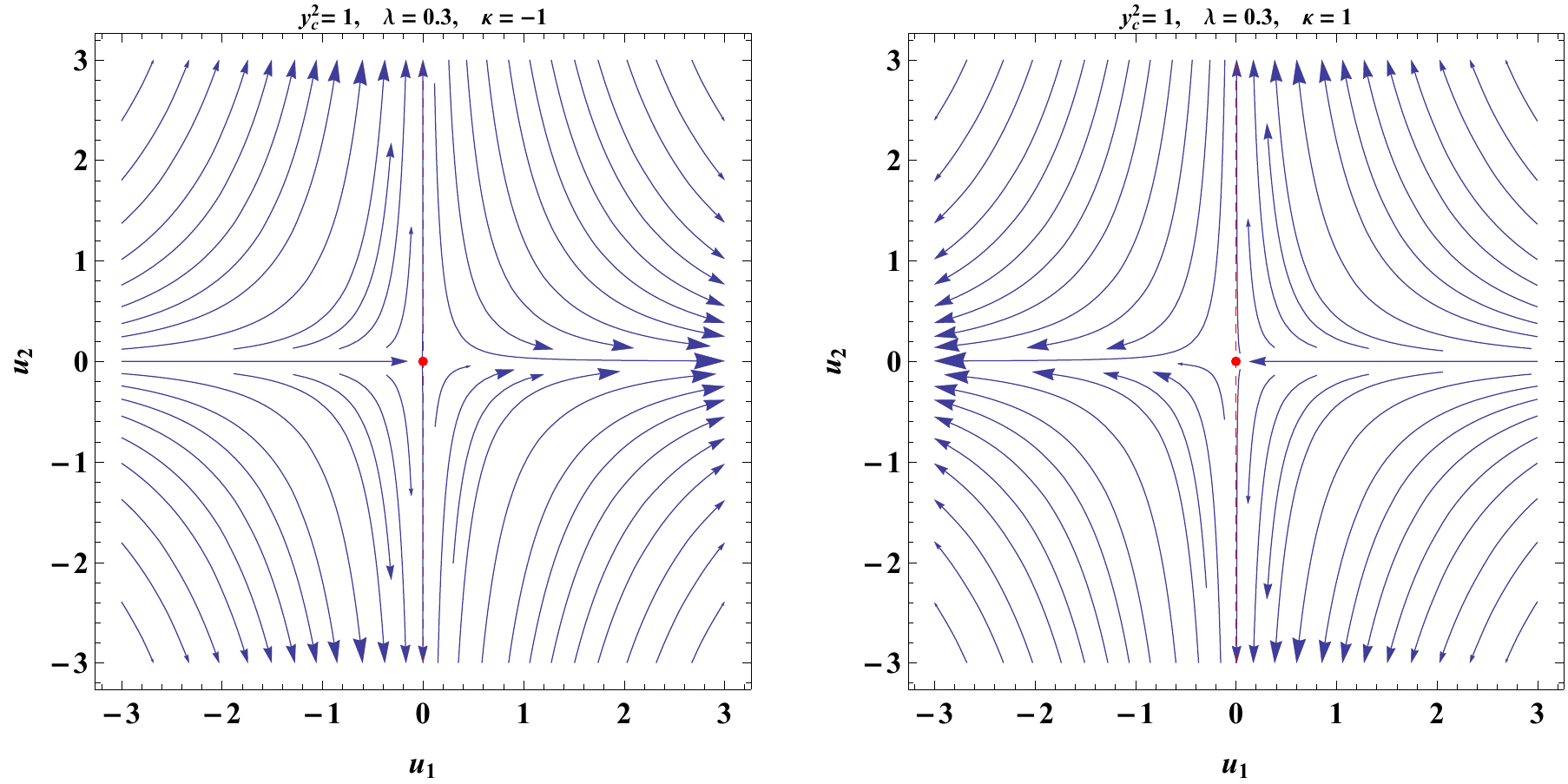}
    \caption{Phase plot of system \eqref{dscenter4} for some values of the parameters showing that $A_{2}^{^{\prime}}$ is a saddle point.}
    \label{fig:my_label4}
\end{figure}
From equation \eqref{centere} the center is unstableto perturbations along the $u_1$-axis. This is due to \eqref{centere} which is a gradient like equation  (saddle behavior) 
\begin{equation}
    u_1'=-U'(u_1), \quad U(u_1)=\frac{1}{72} u_1^2 \left(\sqrt{6} \kappa -6 \lambda \right),
\end{equation}
such that $u_1=0$ is an inflection point of $U(u_1)$. That is, the solution either depart or approach the origin along the $u_1$-axis, depending on the sign of the initial value of $u_1$,  $u_{10}$ and the sign of $\sqrt{6} \kappa -6 \lambda$.

In Fig. \ref{fig:my_label4} a phase plot of system \eqref{dscenter3} is presented for some values of the parameters showing that $A_{2}^{^{\prime}}$ is a saddle point.

\paragraph{Subcase $y_c^2=1$, $2 \alpha -\kappa +\sqrt{6} \lambda =0, \alpha \neq 0$.} 
In this subcase the similarity matrix is
\begin{align}
    s= \left(
\begin{array}{ccccc}
 0 & 0 & 1 & 0 & 0 \\
 \frac{3 \epsilon }{2 \left(\kappa -\sqrt{6} \lambda \right)} & -\frac{\kappa  \epsilon }{2} & 0 & 0 & \frac{\epsilon }{2} \\
 0 & \frac{1}{3} \kappa  \left(\kappa -\sqrt{6} \lambda \right) & 0 & 0 & \sqrt{\frac{2}{3}} \lambda  \\
 0 & 0 & 0 & 0 & 1 \\
 0 & 0 & 0 & 1 & 0 \\
\end{array}
\right)\,,
\end{align}
where $\epsilon=\pm 1$ is the value of $y_c$ and the real Jordan matrix is the same as \eqref{matrix50}.

Defining 
\begin{subequations}
\begin{align*}
& v_1=   \frac{1}{3} \left(3 x_2-\kappa  (-2 y_1 \epsilon +y_2+1)+\sqrt{6} \lambda  (1-2 y_1 \epsilon )\right),\\
& v_2= \frac{\kappa +3 x_2-\sqrt{6} \lambda 
   (y_2+1)}{\kappa  \left(\kappa -\sqrt{6} \lambda \right)}, \; v_3=x_1, \; u_1= \mu, \; u_2= y_2\,,
 \end{align*}
 \end{subequations}
and using the CMT we obtain that the center manifold is given by \eqref{XgraphmanifoldD}, where the $h_i$ satisfy \eqref{pdes}. \newline 
Using the ansatz \eqref{taylor} with $N=2$, we obtain the nonzero coefficients 
$a^{[1]}_{20}= 0, a^{[2]}_{20}=  0, a^{[3]}_{20}=  0, a^{[1]}_{21}=  \frac{1}{\sqrt{6}}-\frac{1}{108} \kappa  \left(\sqrt{6} \kappa -6 \lambda \right), a^{[2]}_{21}=  \frac{-\kappa ^2+6 \lambda ^2+18}{6 \sqrt{6} \kappa ^2-36 \kappa  \lambda
   }, a^{[3]}_{21}=  -\frac{1}{2 \sqrt{6}}, a^{[1]}_{22}=  \frac{1}{12} \left(\sqrt{6} \lambda -\kappa \right), a^{[2]}_{22}=  0, a^{[3]}_{22}=  0$. 
Therefore, the dynamics on the center manifold is governed up to third order by 
\eqref{dscenter4} and we have the same result as before that $A_{2}^{^{\prime}}$ is a saddle point.   
   
\paragraph{Subcase $y_c^2=1$, $\kappa =\sqrt{6} \lambda, \alpha = 0$.}

   \begin{figure}[ptb]
\centering\includegraphics[width=0.8\textwidth]{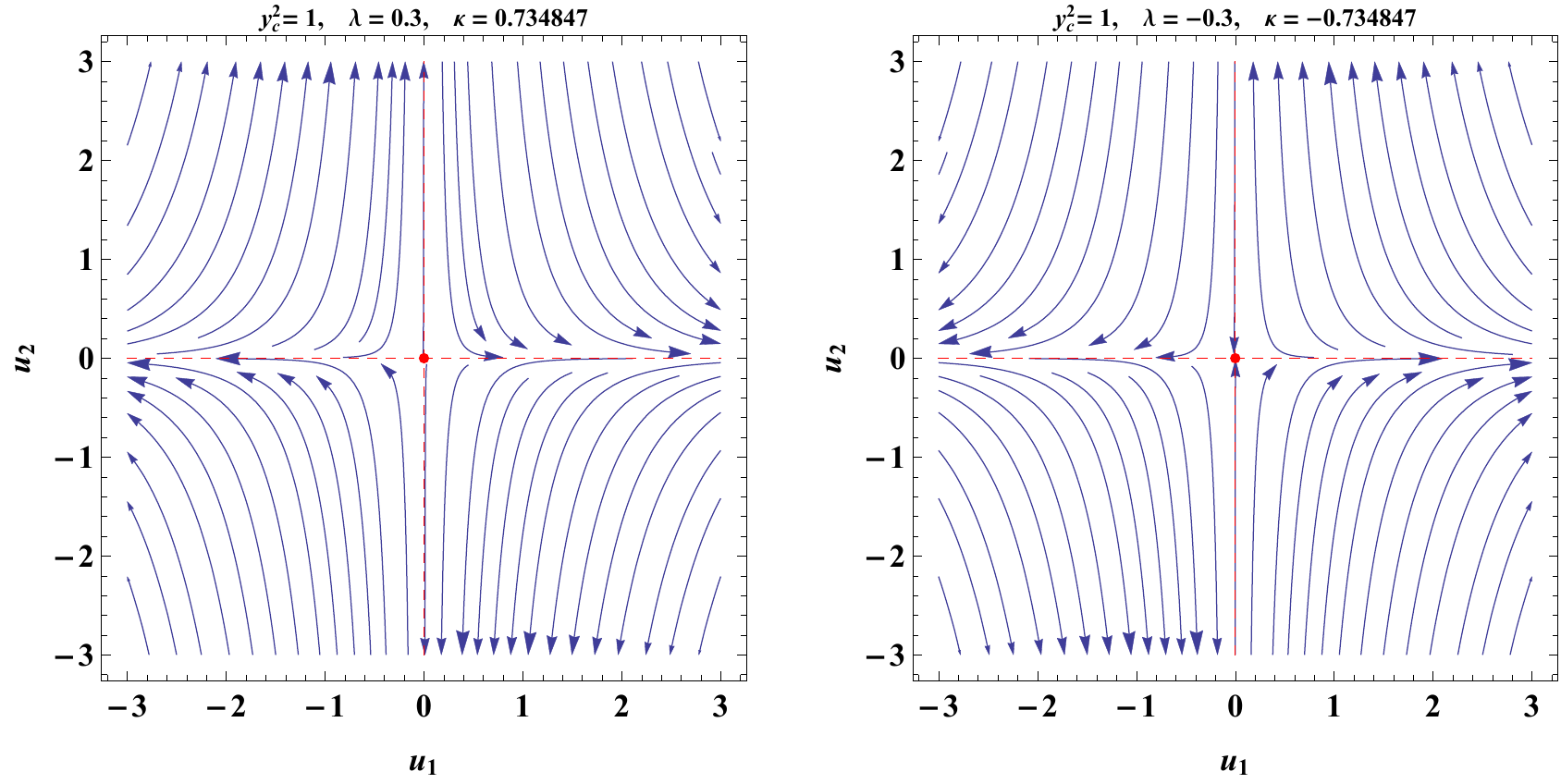}
    \caption{Phase plot of system \eqref{dscenter6} for some values of the parameters showing that $A_{2}^{^{\prime}}$ is a saddle point.}
    \label{fig:my_label6}
\end{figure}

In this subcase the similarity matrix is
\begin{align}
  s=   \left(
\begin{array}{ccccc}
 0 & 0 & -\frac{\sqrt{\frac{2}{3}} \epsilon }{\lambda } & 0 & 0 \\
 0 & 1 & 0 & 0 & \frac{\epsilon }{2} \\
 1 & 0 & 0 & 0 & \sqrt{\frac{2}{3}} \lambda  \\
 0 & 0 & 0 & 0 & 1 \\
 0 & 0 & 0 & 1 & 0 \\
\end{array}
\right)\,,
\end{align}
where $\epsilon=\pm 1$ is the value of $y_c$ and the real Jordan matrix is the same as \eqref{matrix50}.

Defining 
\begin{subequations}
\begin{align*}
& v_1=x_2-\sqrt{\frac{2}{3}} \lambda  y_2,  \; v_2= y_1-\frac{1}{2} (y_2+2) \epsilon, \;  v_3=-\sqrt{\frac{3}{2}} \lambda  x_1 \epsilon, \; u_1=\mu, \;  u_2= y_2 \,,
 \end{align*}
 \end{subequations}
and using the CMT we obtain that the center manifold is given by \eqref{XgraphmanifoldD}, where the $h_i$ satisfy \eqref{pdes}. \newline 
Using the ansatz \eqref{taylor} with $N=3$, we obtain the coefficients 
$a^{[1]}_{20}=0, a^{[2]}_{20}=0, a^{[3]}_{20}= 0, a^{[1]}_{21}=\frac{1}{\sqrt{6}}, a^{[2]}_{21}= \frac{\lambda  \epsilon }{12},  a^{[3]}_{21}=\frac{\lambda  \epsilon }{4},  a^{[1]}_{22}= 0, a^{[2]}_{22}= -\frac{\epsilon }{8}, a^{[1]}_{22}= 0, a^{[1]}_{30}=0, a^{[2]}_{30}= 0, a^{[3]}_{30}= 0, a^{[1]}_{31}= 0, a^{[2]}_{31}=0, a^{[3]}_{31}= 0, a^{[1]}_{32}= \frac{5 \lambda ^2}{6 \sqrt{6}},  a^{[2]}_{32}= \frac{\lambda  \epsilon }{24},  a^{[3]}_{32}=0, a^{[1]}_{33}=0, a^{[2]}_{33}=\frac{\epsilon ^3}{16}, a^{[3]}_{33}= 0$. 

Therefore, the dynamics on the center manifold is governed up to fourth  order by 
\begin{align}
\label{dscenter6} 
& u_1'= \frac{1}{4} \lambda  u_1^2 u_2, \; 
u_2'=- \lambda  u_1 u_2^2.
\end{align}
In Fig. \ref{fig:my_label6} a phase plot of system \eqref{dscenter6} is presented for some values of the parameters showing that $A_{2}^{^{\prime}}$ is a saddle point.
\section{Cosmological linear perturbations}

\label{sec5}

We continue our analysis by investigating the modified scalar field Lagrangian's effects in the cosmological linear perturbation theory. For this purpose, we consider the simple linear perturbation theory in the Newtonian gauge, where the spacetime line element is described as \cite{amebook}
\begin{equation}
ds^{2}=a^{2}\left(  \eta\right)  \left(  -\left(  1+2\Phi\left(
\eta,x,y,z\right)  \right)  d\eta^{2}+\left(  1-2\Phi\left(  \eta
,x,y,z\right)  \right)  \left(  dx^{2}+dy^{2}+dz^{2}\right)  \right),
\label{pr.01}%
\end{equation}
where $\eta$ ~($= \int a^{-1} dt$) is the conformal time  . 

Before presenting the perturbed equations, let us define the perturbed physical quantities of the usual scalar field $\phi$, the scalar field $\psi$ due to the effect of GUP, the matter-energy density, the DE density, the DE pressure as follows:
\begin{gather}
\phi+\delta \phi, ~~~\psi +\delta\psi,~~~\rho_m+\delta \rho_m,~~~\rho_d+\delta \rho_d,~~~p_d+\delta p_d\,.
\end{gather}
We note here that $\phi, \psi, \rho_m, \rho_d, p_d$ denote the corresponding background quantities.

For the line element (\ref{pr.01}), the perturbed terms of the Einstein tensor
are \cite{amebook}%
\begin{align}
\delta G_{0}^{0}  &  =-\frac{2}{a^{2}}\nabla^{2}\Phi+\frac{6}{a}%
\mathcal{H}\left(  \dot{\Phi}+a\mathcal{H}\Phi\right)\,, \label{pr.02}\\
\delta G_{i}^{0}  &  =-\frac{2}{a^{2}}\nabla_{i}\left(  \dot{\Phi
}+a\mathcal{H}\Phi\right)\,, \label{pr.03}\\
\delta G_{j}^{i}  &  =2\Phi\left(  \frac{2}{a}\mathcal{\dot{H}}+3\mathcal{H}%
^{2}\right)  +\frac{2}{a^{2}}\left(  \ddot{\Phi}+3a\mathcal{H}\dot{\Phi
}\right) \,, \label{pr.04}%
\end{align}
where now dot means the derivative with respect to the variable $\eta$, and
$\mathcal{H}=\frac{\dot{a}}{a^{2}}$ is the Hubble parameter in the new frame.

We assume the comoving observer $u^{\mu}=\left(  \frac{1-\Phi}{a},\delta
u^{i}\right)  $ such that the energy momentum tensor for the DM to
have nonzero perturbed components \cite{amebook}%
\begin{equation}
\delta T_{0}^{\left(  m\right)  0}=-\rho_{m}\delta_{m}~,~\delta T_{0}%
^{i}=a\rho_{m}\delta u^{i}~\text{,} \label{pr.05}%
\end{equation}
where $\delta_{m}=\delta\rho_{m}/\rho_{m}$. Similarly for the scalar
field it follows%
\begin{equation}
\delta T_{0}^{\left(  d\right)  0}=-\frac{1}{a^{2}}\left(  \dot{\phi}
\dot{\left(\delta\phi\right)}  -\dot{\phi}^{2}\Phi\right)  -V_{,\phi}\delta
\phi-\frac{2\beta\hbar^{2}}{a^{2}}\left( \dot{\left(\delta\phi\right)}
\dot{\psi}-2\dot{\phi}\dot{\psi}\Phi\right)  -\frac{2\beta\hbar^{2}}{a^{2}%
}\dot{\phi}   \dot{\left(\delta\psi\right)}  +2\beta\hbar^{2}\psi\delta\psi,
\label{pr.06}%
\end{equation}

\begin{equation}
\delta T_{i}^{\left(  d\right)  i}=\frac{1}{a^{2}}\left(  \dot{\phi}\dot{\left(\delta\phi\right)}   -\dot{\phi}^{2}\Phi\right)  -V_{,\phi}\delta
\phi+\frac{2\beta\hbar^{2}}{a^{2}}\left(  \dot{\left(\delta\phi\right)} 
\dot{\psi}-2\dot{\phi}\dot{\psi}\Phi\right)  +\frac{2\beta\hbar^{2}}{a^{2}%
}\dot{\phi}   \dot{\left(\delta\psi\right)}  +2\beta\hbar^{2}\psi\delta\psi,
\label{pr.07}%
\end{equation}%
\begin{equation}
\delta T_{i}^{\left(  d\right)  0}=-\frac{\dot{\phi}}{a^{2}}\left(  \delta
\phi+4\beta\hbar^{2}\delta\psi\right)  _{,i}, \label{pr.08}%
\end{equation}
where we have replaced $\phi\rightarrow\phi+\delta\phi$ and $\psi
\rightarrow\psi+\delta\psi$. \ However, $\delta\phi~$and $\delta\psi$ are not independent, they are related with the constraint condition (\ref{gp.21}). At
this point, someone could omit the terms $\beta\hbar^{2}\delta\phi$%
,$~\beta\hbar^{2}\delta\psi$, etc., because they are small. However, it is
important to remark that in terms of $\beta\hbar^{2}$, the
related dynamical system is a singular perturbation system, and as it has been shown before in \cite{gup1}  that the singular perturbation term can be relatively large such that it can drive the dynamics in the background space. That is why we found the
existence of de Sitter asymptotic universe for the background space in  the previous section.

We now proceed with our analysis by considering the interacting model of previous
section with $\kappa=0$, that is $Q_{A}=\frac{\alpha}{\sqrt{6}a}\left(
\dot{\phi}\delta_{m}+\rho_{m}\delta\dot{\phi}\right)  $. Therefore, the
perturbative equations for the matter components are%
\begin{equation}
\dot{\delta}_{m}-3\dot{\Phi}+a\left(  \delta u^{i}\right)  _{,i}-\frac{\alpha
}{\sqrt{6}}\left(  \Phi\dot{\phi}+\delta\dot{\phi}\right)  =0, \label{pr.09}%
\end{equation}
where we have replaced $Q=Q_{0}+\delta Q$, in which $Q_{0}$ is the unperturbed
term and $\delta Q$ is the linear perturbation term.

For the scalar field it follows%
\begin{equation}
\frac{1}{a^{2}}\left(  \ddot{(\delta\phi)}-\nabla^{2}\left(  \delta\phi\right)
-4\dot{\Phi}\dot{\phi}\right)  +\frac{2}{a}\mathcal{H}\delta\dot{\phi}%
-2\Phi\psi-\delta\psi=0, \label{pr.10}%
\end{equation}%
\begin{equation}
\frac{2\beta\hbar^{2}}{a^{2}}\left(  \left(  \delta\ddot{\psi}\right)
-\nabla^{2}\left(  \delta\psi\right)  -\dot{\psi}\dot{\Phi}+2a\mathcal{H}%
\dot{\left(\delta\psi\right)}+2a\mathcal{H}\dot{\psi}\Phi\right)  +\left(  \delta
\psi+V_{,\phi\phi}\delta\phi\right)  +2\left(  \psi+V_{,\phi}\right)
\Phi+\frac{2\alpha}{\sqrt{6}}\Phi\rho_{m}+\frac{\alpha}{\sqrt{6}}\rho
_{m}\delta_{m}=0\,, \label{pr.11}%
\end{equation}
where we have omitted the presentation of the equations for the velocity divergence.

From equations (\ref{pr.10}) and (\ref{pr.11}) it is clear that for
$\beta\hbar^{2}=0$, the equations for  quintessence are recovered
\cite{am0}. However, \ from equation (\ref{pr.11}), we observe that it is
singular in terms of $\beta\hbar^{2},$ which means that the singular
pertubative term $\left(  \left(  \delta\ddot{\psi}\right)  -\nabla^{2}\left(
\delta\psi\right)  +\frac{2}{a}\mathcal{H}\dot{\psi}\Phi-4\dot{\Phi}\dot{\psi
}\right)  $ can be significantly large such that it can dominate and drive the dynamics.

We continue by define the new independent variable to be $\tau=\ln a$, that is
$\frac{1}{a}\frac{d}{d\eta}=\mathcal{H}\frac{d}{d\ln a}$.  Hence, by writing in Fourier form and
setting $\delta u^{i}=0$, for the exponential potential, the pertubative equations
(\ref{pr.09}), (\ref{pr.10}) and (\ref{pr.11}) are written as follows%
\begin{equation}
0=\text{$\delta$}_{m}^{\prime}-3\Phi^{\prime}-\frac{\alpha}{\sqrt{6}}\left(
\delta\phi^{\prime}+2\sqrt{6}x_{1}\Phi\right)  , \label{pe.01}%
\end{equation}

\begin{equation}
0=\delta\phi^{\prime\prime}+\left(  \frac
{\mathcal{H}^{\prime}}{\mathcal{H}}+4\right)  \delta\phi^{\prime}-4\sqrt
{6}x_{1}\Phi^{\prime}  +\frac{k^{2}}{a^2 \mathcal{H}^{2}}\delta\phi - \frac{\delta\psi+3\mu y_{2}\Phi
}{\mathcal{H}^{2}}, \label{pe.02}%
\end{equation}

\begin{equation}
0=2\beta\hbar^{2}\left(  \delta\psi^{\prime\prime}+\left(  \frac
{\mathcal{H}^{\prime}}{\mathcal{H}}+4\right)  \delta\psi^{\prime}+\frac{k^{2}%
}{a^{2}\mathcal{H}^{2}}\delta\psi\right)  +\frac{\delta\psi+3\mu y_{2}\Phi
}{\mathcal{H}^{2}}+3y_{1}^{2}\lambda\left(  \lambda\delta\phi-2\Phi\right)
+2\sqrt{6}x_{2}\left(  \Phi-\Phi^{\prime}\right)  +\frac{3}{\sqrt{6}}%
\alpha\left(  \delta_{m}+2\Phi\right), \label{pe.03}%
\end{equation}
where a prime means
derivative with respect to $\tau=\ln a.$ 
Recalling
\begin{align}
  \frac
{\mathcal{H}^{\prime}}{\mathcal{H}}= -\frac{3}{2}(1+ w_{\rm eff}), 
\end{align}
where $w_{\rm eff}$ is given by \eqref{gp.29}, one could write,
\begin{equation}
  \frac
{\mathcal{H}^{\prime}}{\mathcal{H}}+4=    \frac{1}{2} \left(5-3 x_1 (x_1+x_2)+3 y_1^2-3
   y_2\right).
\end{equation}

Note that the above equations
can be simplified by considering the Poisson equation,
which in sub-horizon scales ($k/a\gg \mathcal{H}$) becomes \cite{Ma:1995ey}:
\begin{equation}
    -\frac{k^2}{a^2 \mathcal{H}^2 }\Phi = \frac{3}{2} \left(\Omega_m \delta_m + (1+ 3 c_{d}^2)\Omega_{d} \delta_{d}\right),
\end{equation}
where  $c_{d}^2$ is the effective sound speed of DE
perturbations (the corresponding quantity for matter is
zero in the dust case), 
\begin{align}
  &  c_{d}^2 = w_{d} - \frac{w_{d}'}{3(1+w_{d})},
    \\
  &  \Omega_{d}= \left(  x_{1}^{2}+y_{1}^{2}\right)  + \left(  x_{1}x_{2}%
-y_{2}\right), \\
& \Omega_{m}=1-\left(  x_{1}^{2}+y_{1}^{2}\right)  -\left(  x_{1}x_{2}%
-y_{2}\right),\\
& \Omega_{d} \delta_{d} \equiv - \Omega_{d}\delta T_{0}^{\left(  d\right)  0}/ \rho_{d}= -\frac{\delta T_{0}^{\left(  d\right)  0}}{3 \mathcal{H}^2},
\end{align}
where we have used 
$\delta T_{0}^{\left(  d\right)  0}=-\rho_{d}\delta_{d}$ with 
$\delta T_{0}^{\left(  d\right)  0}$ given by 
\eqref{pr.06}.
That is, 
\begin{align}
\Omega_{d} \delta_{d}=    \frac{1}{3 \mathcal{H}^2 a^{2}}\left(  \dot{\phi}
\dot{\left(\delta\phi\right)}  -\dot{\phi}^{2}\Phi\right)  +\frac{V_{,\phi}\delta
\phi}{3 \mathcal{H}^2}+\frac{2\beta\hbar^{2}}{3 \mathcal{H}^2 a^{2}}\left( \dot{\left(\delta\phi\right)}
\dot{\psi}-2\dot{\phi}\dot{\psi}\Phi\right)  +\frac{2\beta\hbar^{2}}{3 \mathcal{H}^2 a^{2}%
}\dot{\phi}   \dot{\left(\delta\psi\right)}  -2\beta\hbar^{2} \frac{\psi\delta\psi}{3 \mathcal{H}^2}\,.
\end{align}
Using 
$\dot{\left(\delta\phi\right)}= a \mathcal{H} {\delta\phi}', \dot{\phi}= a \mathcal{H} \phi', \dot{\left(\delta\psi\right)}= a \mathcal{H} {\delta\psi}', \dot{\psi}= a \mathcal{H} \psi'$, where the dot denotes derivative with respect to  the conformal time $\eta$, and $\phi '(\tau )=\sqrt{6} x_1, V(\phi )=3 y_1^2 \mathcal{H}^2, \psi '(\tau
   )=\frac{\sqrt{\frac{3}{2}} x_2}{2 \beta \hbar^2},\psi = -\frac{3}{2} \mathcal{H}^2 \mu y_2$, we obtain 
\begin{align}
\Omega_{d} \delta_{d} & =    \left(  {\phi}'
{\delta\phi}'  -{\phi'}^{2}\Phi\right) +\frac{2\beta\hbar^{2}}{3  }\left( {\delta\phi}' {\psi}'-2 {\phi}' {\psi}' \Phi\right)  +\frac{2\beta\hbar^{2}}{3 %
} {\phi}'    { \delta\psi}'  +\frac{V_{,\phi}\delta
\phi}{3 \mathcal{H}^2} -2\beta\hbar^{2} \frac{\psi\delta\psi}{3 \mathcal{H}^2}\,, \nonumber \\
&  =   \left(  {\phi}'
{\delta\phi}'  -{\phi'}^{2}\Phi\right) +\frac{2\beta\hbar^{2}}{3  }\left( {\delta\phi}' {\psi}'-2 {\phi}' {\psi}' \Phi\right)  +\frac{2\beta\hbar^{2}}{3 %
} {\phi}'    { \delta\psi}'  -\lambda \frac{V\delta
\phi}{3 \mathcal{H}^2} -2\beta\hbar^{2} \frac{\psi\delta\psi}{3 \mathcal{H}^2}\,, \nonumber \\
& = - 2 \Phi  x_1^2 (4 \beta  \hbar^2 +3) +\ 2 \sqrt{\frac{2}{3}} \beta 
   x_1 \hbar^2  {\delta\psi}' +\frac{(6 x_1+x_2) \delta \phi '}{\sqrt{6}}-\lambda  y_1^2 {\delta\phi}+\beta  \mu  y_2 \hbar^2  {\delta\psi}\,.
\end{align}

The  effective sound speed of the DE
perturbations $c_{d}^2$ for $\kappa=0$ can be expressed as 
\begin{align}
& c_{d}^2=   \frac{x_1 x_2^2 \left(12 x_1-\sqrt{6} \mu  y_2\right)+x_2 y_2 \left(12 x_1+\sqrt{6} \mu  y_2\right)-2 \sqrt{6} x_1 y_2 \left(2 \lambda  y_1^2+\mu 
   y_2\right)}{12 x_1 x_2 (x_1 x_2+y_2)} \nonumber \\
   & -\frac{\alpha  y_2 \left(x_1 (x_1+x_2)+y_1^2-y_2-1\right)}{3 x_2 (x_1 x_2+y_2)}. 
\end{align}
 In what follows, we are going to explore how scalar perturbations behave in the  matter dominated  solutions obtained from the background analysis.

\subsection{Matter era}

Consider now that we are near the stationary point $A_{3}$, then from the
linear perturbation terms of the Einstein field equations and (\ref{pe.01}%
)-(\ref{pe.03}) it follows%

\begin{align}
&\delta_{m}^{\prime\prime}+\frac{7}{2}\delta_{m}^{\prime}+\frac{\alpha}%
{\sqrt{6}}\left(  2\delta\phi^{\prime}-\delta_{\psi}\right)  
=0,\label{pe.04}\\
&\delta\phi^{\prime\prime}+\left(  \frac{\mathcal{H}^{\prime}}{\mathcal{H}%
}+4\right)  \delta\phi^{\prime}-\delta_{\psi}    =0\,,\label{pe.05}\\
&2\beta\hbar^{2}\mathcal{H}^{2}\left(  \delta_{\psi}^{\prime\prime}+\left(
\frac{\mathcal{H}^{\prime}}{\mathcal{H}}+4\right)  \delta\psi^{\prime}%
+\frac{33}{2}\delta_{\psi}\right)  +\delta_{\psi}+\frac{2\alpha}{3\sqrt{6}%
}\left(  6\delta_{m}+5\delta_{m}^{\prime}-6\zeta\delta\phi^{\prime}\right)
  =0, \label{pe.06}%
\end{align}
in which $\delta\psi=\mathcal{H}^{2}\delta_{\psi}$, $\mathcal{H}^{2}\left(
\tau\right)  =\mathcal{H}_{0}^{2}e^{-3\tau}$ and for simplicity, we have
assumed $k=0$. The latter system is a singular perturbation system that possesses a slow invariant manifold. Therefore, we continue our analysis by studying the evolution in the slow-fast manifolds
\cite{Tikhonov,Fusco}.

Now in the fast manifold we do the change of independent variable $\tau
=2\beta\hbar^{2}s,$ in equation (\ref{pe.06}) the dominated terms are $\left(
\frac{\partial^{2}\delta_{\psi}}{\partial s^{2}}-\frac{3}{2}\frac
{\partial\delta_{\psi}}{\partial s}\right)  \simeq0,~$which provides
$\delta_{\psi}\left(  a\right)  =\delta_{\psi}^{0}+\delta_{\psi}^{1}%
e^{\frac{3}{2}s}$, recall that $\ln a=2\beta\hbar^{2}s$ and we assume that
$\delta_{\psi}^{1}$ is large enough. Therefore, in the fast manifold from
(\ref{pe.04}), (\ref{pe.05}) we find%
\begin{equation}
\frac{\partial^{2}\delta\phi}{\partial s^{2}}-\frac{3}{2}\frac{\partial
\delta\phi}{\partial s}-\delta_{\psi}^{1}e^{\frac{3}{2}s}\simeq0,
\end{equation}
and%
\begin{equation}
\frac{\partial^{2}\delta_{m}}{\partial s^{2}}-\frac{\alpha}{\sqrt{6}}\left(
\bar{\delta}_{\psi}^{1}e^{\frac{3}{2}s}\right)  =0.
\end{equation}

\begin{figure}[ptb]
\centering\includegraphics[width=1\textwidth]{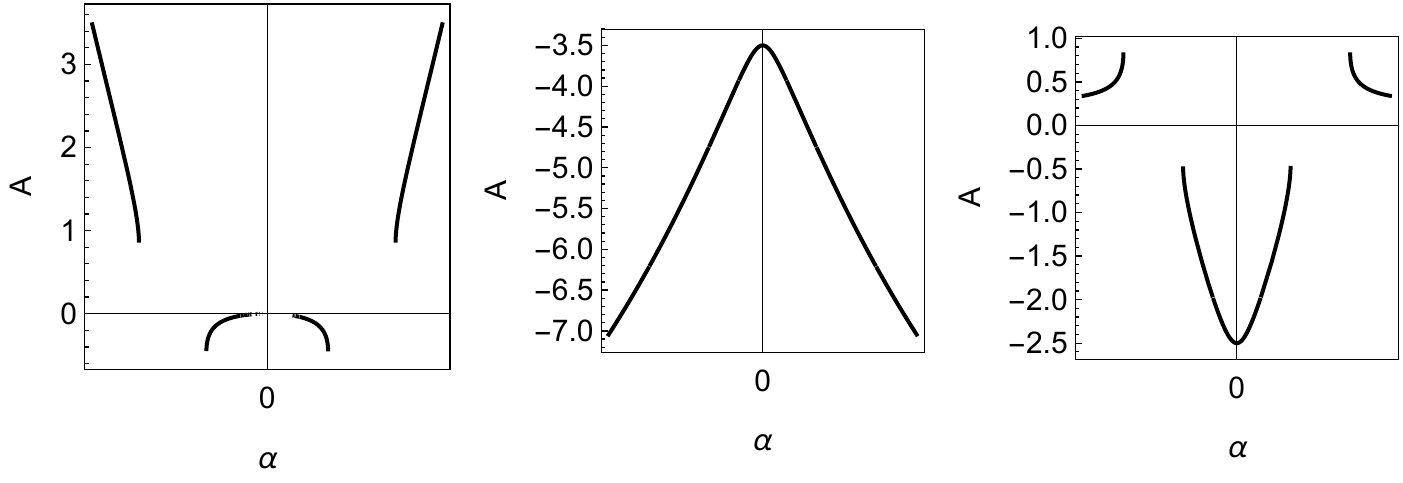} \caption{The dependency of the exponents of
the linear system (\ref{pe.09}), (\ref{pe.10}) with the coupling parameter $\alpha$ is displayed in the plots. }%
\label{fig2A}%
\end{figure}

\begin{figure}[ptb]
\centering\includegraphics[width=6.5cm, height=6.5cm]{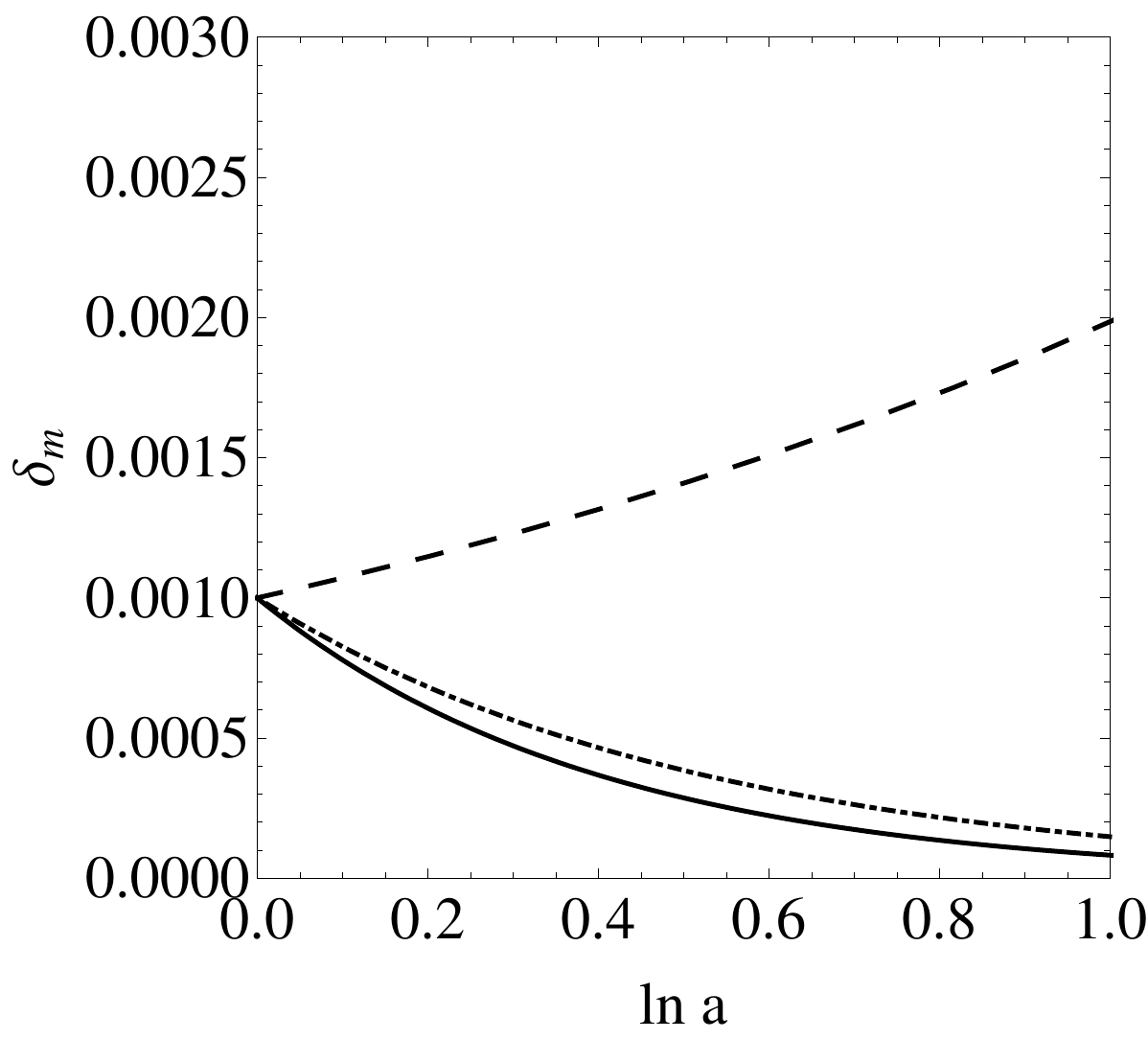} \caption{The evolution of matter density perturbation for $\alpha=0$ (solid curve), $\alpha=1$ (dotted-dashed curve) and $\alpha=3$ (dashed curve).}%
\label{fig3}%
\end{figure}

Therefore,
\begin{align}
\delta\phi\left(  s\right)   &  =\frac{2}{9}e^{\frac{3}{2}s}\left(  \left(
3s-2\right)  \delta_{\psi}^{1}+\delta\phi_{1}\right)  +\delta\phi_{2},\\
\delta_{m}\left(  s\right)   &  =\frac{4}{9}\frac{\alpha}{\sqrt{6}}\left(
\bar{\delta}_{\psi}^{1}e^{\frac{3}{2}s}\right)  +s\delta_{m}^{0}+\delta
_{m}^{0}.
\end{align}
from where we observe that a growing mode exists in the fast manifold because of the higher-order derivative terms of the GUP. Hence, we see that the GUP component's presence enhances scalar perturbations' growth compared to the usual quintessence case.

On the other hand, in the slow manifold where $2\beta\hbar^{2}\rightarrow0$,
we find the system%
\begin{align}
\delta_{m}^{\prime\prime}+\frac{\alpha}{9}\left(  6\alpha\delta_{m}%
+5\alpha\delta_{m}^{\prime}-\sqrt{6}\left(  \alpha^{2}-3\right)  \delta
\phi^{\prime}\right)  +\frac{7}{2}\delta_{m}  &  =0,\label{pe.09}\\
\delta\phi^{\prime\prime}+\frac{\alpha}{9}\left(  \sqrt{6}\left(  6\delta
_{m}+5\delta_{m}^{\prime}\right)  -6\alpha\delta\phi^{\prime}\right)
+\frac{5}{2}\delta\phi^{\prime}  &  =0. \label{pe.10}%
\end{align}
We replace $\delta_{m}\left(  \tau\right)  =\delta_{m}^{0}e^{A\tau}$ and
$\delta\phi\left(  \tau\right)  =\delta\phi_{0}e^{A\tau}$ and the values of
the exponent $A$ depend of the coupling parameter $\alpha$, as they are
presented in Fig. \ref{fig2A}. The behavior of $A$ concerning coupling parameter $\alpha$ exhibits two interesting results. First, the evolution of perturbations is independent of the sign of $\alpha$. Second, unlike the usual quintessence model, there is a decay of scalar perturbations (i.e., $A<0$) in the present model for the uncoupled scenario.  However, scalar perturbations either decay or grow or describe an oscillatory solution for the coupling case. It is worth mentioning that in the case where scalar perturbations decay, the decay rate is slower compared to the uncoupled case. From Fig. \ref{fig3}, it is interesting to see that while for $\alpha=0,~1$ there is a decay of scalar perturbations, however, for $\alpha=3$ there is a growth of scalar perturbations. On assuming a negligible oscillating behavior of $\delta \phi$, the results reduce to that of \cite{am0}. However, in general, such an assumption cannot be applied as $\delta\phi$ may have non-oscillating terms since it is a singular perturbation system.

We have presented a qualitative analysis for the evolution of the perturbations in the slow-fast manifolds. From the above results, it is clear that the modification of the scalar field Action Integral by the GUP indeed modifies the evolution at the perturbation level.

\section{Conclusions}
\label{sec6}

The Heisenberg Uncertainty Principle needs to be modified whenever we consider gravitational interaction. The GUP, which emerges from one such modification, has given rise to a host of exciting results. For instance, the quantum effects due to GUP modifies Hawking temperature and the Bekenstein entropy. Consequently, we get a modified Bekenstein system that critically affects Black hole evaporation. Similarly, in cosmology, the GUP is also very effective in explaining various issues related to the universe's evolution.  The cosmological background dynamics of the GUP modified quintessence scalar field analyzed in  \cite{gup2} provides engaging scenarios which are absent in the usual quintessence scalar field model. These exciting results motivate us to extend the work of \cite{gup2} to the  GUP modified interacting quintessence scenario at the background and perturbation levels.

More elaborately, here, we investigated a very generalized cosmological scenario of recent interest. We consider a quintessence scalar field cosmological model in a spatially flat FLRW background space in which the Lagrangian of the scalar field has been modified according to the quadratic GUP.  Besides, we assumed that the scalar field is coupled to the DM component of the cosmological fluid. We considered three models already proposed in the literature for the interacting functions between the two fluid sources. For each cosmological model, we investigated the evolution of the global dynamics and studied the asymptotic behaviour.

First, we analyze the interacting DE model's dynamics at the background level using the dynamical analysis tools.  Indeed, we found that the models' cosmological evolution is different in the presence of the minimum length or its absence.  One of the present model's interesting features compared to the corresponding interacting model of usual quintessence is that it exhibits matter-dominated and de Sitter solutions. While the former is crucial to explain the structure formation, the latter describes the late time acceleration at the background level.   It is worth mentioning that these two solutions are due to the  GUP modification and are independent of the choice of potential or coupling functions.  Unlike the usual quintessence theory, the GUP modified interacting model cannot provide a solution that alleviates the coincidence problem.

To better understand background dynamics, we explored the model's behavior at the linear perturbation level. More precisely, we focused on the scalar perturbations.  We derived the dynamical equations for linear cosmological perturbation, where we found that the resulting linear equations form a singular perturbation system. Therefore, in contrast to the usual quintessence,   we can explicitly write the perturbed equations' solution in the fast and slow manifolds. In the fast manifold, the  GUP components enhance the scalar perturbations' growth due to the higher-order derivative terms of the GUP. However,  in the slow manifold, the scalar perturbations either decay or grow or describe an oscillatory solution.  Consequently, in the presence of the minimum length, the perturbation equations are also affected. It means that the present model differs from the usual quintessence scalar field theory.

In Ref. \cite{cgup1}, the possibility to solve the
cosmological constant problem with the introduction of the minimum length and
specifically with the quadratic GUP has been investigated. However, the predicted value of the
cosmological constant has been found to be $\Lambda\sim\frac{1}{\beta^{2}}%
$ which cannot solve the smallness problem for the cosmological
constant. A similar result was found recently in \cite{cgup2} by introducing a
more general expression for the GUP where the authors found that%
$\Lambda\sim\frac{1}{\beta^{4}}$. On the other hand, quintessence has
been introduced as a theoretical model to introduce a time-varying effects of
the DE fluid. However, for the simple exponential potential for the
quintessence model the de Sitter universe is not recovered
\cite{Copeland:2006wr}. Thus, this is not the case where the minimum length is
introduced to modify the scalar field Lagrangian. As we investigated in this
study, the presence of the minimum length \ modifies the field equations in such a way so that 
the quintessence may reach the limit of the cosmological constant. This is an
interesting mechanism in order the field equation to reach the de Sitter
limit. The de Sitter expansion era can solve the  \textquotedblleft
flatness\textquotedblright, \textquotedblleft horizon\textquotedblright\ and 
monopole problems \cite{f1,f2} as it is explained by the support the cosmic
\textquotedblleft no-hair\textquotedblright\ conjecture \cite{nh1,nh2}.
Furthermore, because the deformation parameter $\beta$ is found to be
multiplied always with the derivatives of the scalar field, the new degrees of
freedom can solve the smallness problem of the cosmological constant. For
instance, considering the de Sitter point $A_{2}$of the interaction
model~$Q_{A}$, it follows%
\[
\sqrt{\frac{2}{3}}\lambda y_{1}^{2}=\beta\hbar^{2}\frac{2\sqrt{2}\dot{\psi}%
}{\sqrt{3}H_{0}}~~,~\sqrt{3}y_{1}=\frac{\sqrt{V}}{3H_{0}}~,~y_{1}^{2}%
-1=\frac{\beta\hbar^{2}\psi^{2}}{3H_{0}^{2}}\,\ \text{with }H_{0}=\sqrt
{\frac{2}{3}\Lambda}%
\]
Hence, it is clear that $\Lambda$ does not depend explicitly
on $\beta.$ However, if we assume $\lambda=0$ and $y_{1}%
=0$, \ which means that the initial scalar field model is the stiff
fluid, from the latter expressions, it follows $\Lambda=-\frac{\beta\hbar
^{2}\psi^{2}}{2}$. Hence, either from a simple quintessence model
which does not provide any inflationary era, the modification of the
Lagrangian function by the GUP provides an inflationary epoch driven by the
higher-order derivatives of the scalar field.

This work contributes to the study of the existence of the minimum length in cosmological models.  It would be interesting to investigate the constraint of the model concerning the upcoming precise cosmological observations in future works.

\begin{acknowledgments}

This work is based on the research supported in part by the National Research Foundation of South Africa (Grant Number 131604). The research of AP and GL was funded by Agencia Nacional de Investigaci\'{o}n
y Desarrollo - ANID through the program FONDECYT Iniciaci\'{o}n grant no.
11180126. Additionally, GL was also funded by Vicerrector\'{\i}a de
Investigaci\'{o}n y Desarrollo Tecnol\'{o}gico at Universidad Cat\'{o}lica del Norte. JD was supported by the Core Research Grant of SERB, Department of Science and Technology India (File No. CRG $\slash 2018 \slash 001035$) and the Associate program of IUCAA. SP has been supported by the Mathematical Research Impact-Centric Support Scheme  (MATRICS), File No. MTR/2018/000940, given by the Science and Engineering Research Board (SERB), Govt. of India. 

\end{acknowledgments}

\end{document}